\documentclass[11pt,a4paper]{article}

\usepackage{latexsym}
\usepackage{amsmath}
\usepackage{amsfonts}
\usepackage{amsbsy}
\usepackage{amssymb}
\usepackage{epsfig}
\usepackage{mathrsfs}
\usepackage{epsfig}
\usepackage{psfrag}
\usepackage{bbm}
\usepackage{graphicx,color}
\usepackage{hyperref}

\numberwithin{equation}{section}

\def\be{\begin{equation}}
\def\ee{\end{equation}}
\def\ba{\begin{eqnarray}}
\def\ea{\end{eqnarray}}

\newcommand\nn{\nonumber}
\newcommand\q{\quad}

\title{Path integral measure and triangulation independence in discrete gravity}
\author{Bianca Dittrich, Sebastian Steinhaus\\[5pt]
\small   MPI for Gravitational Physics,\\
 \small Am M\"uhlenberg 1, D-14476 Potsdam, Germany }

 \date{}

\begin{document}

\maketitle

\begin{abstract}

A path integral measure for gravity should also preserve the fundamental symmetry of general relativity, which is diffeomorphism symmetry. In previous work, we argued  that a successful implementation of this symmetry into discrete quantum gravity models would imply discretization independence. We therefore consider the requirement of triangulation independence for the measure in (linearized) Regge calculus, which is a discrete model for quantum gravity, appearing in the semi--classical limit of spin foam models. To this end we develop a technique to evaluate the  linearized Regge action associated to Pachner moves in 3D and 4D and show that it has  a simple, factorized structure. We succeed in finding a local measure for 3D (linearized) Regge calculus that leads to triangulation independence. This measure factor coincides with the asymptotics of the Ponzano Regge Model, a 3D spin foam model for gravity. We furthermore discuss to which extent one can find a triangulation independent measure for 4D Regge calculus and how such a measure would be related to a quantum model for 4D flat space. To this end, we also determine the dependence of classical Regge calculus on the choice of triangulation in 3D and 4D.

\end{abstract}


\section{Introduction}

Many approaches to quantum gravity, such as spin foams \cite{spinfoams}, group field theories \cite{gft}, (causal) dynamical triangulations \cite{edt,cdt} and Regge quantum gravity \cite{regge}, rely on a path integral approach. A (non--perturbative) path integral has to be regularized to make it well defined. In the process of this regularization, several choices have to be made, that differ in the various approaches. Broadly one can understand these choices as deciding on a measure on the space of all geometries. This includes various aspects, such as to define the space of geometries, for example the space of all triangulations with fixed edge lengths in dynamical triangulations versus the space defined by allowing all possible edge lengths (satisfying generalized triangle inequalities) in a fixed triangulation such as in Regge calculus, or some generalized discrete geometric spaces, as appearing in loop quantum gravity \cite{dittrichryan, twisted}.  A related question is whether to include a sum over  triangulations, such as in (causal) dynamical triangulations and group field theories, or even over two--complexes as suggested for spin foam models \cite{ben_knotting}. Alternatively the path integral may just include an integration over geometric data associated to a given, fixed, discretization. For discussions on the relation between these approaches, see \cite{freidelgft,dittrichreview,matteo}.

One reason for this many different suggestions, is that the space of all (discrete) geometries and its relation to the corresponding continuum space needs to be better understood \cite{carforabook,lollreview}.  Many difficulties are rooted in the role of diffeomorphism symmetry, by which the space of  metrics has to be quotiened to obtain the space of geometries. Discretizations obscure the role of diffeomorphisms, see \cite{dittrichreview,bahrreview} for a discussion. In particular, for a precise notion of diffeomorphism symmetry in the discrete \cite{dittrichreview}, one can show that this symmetry is broken for 4D Regge gravity \cite{bahrdittrich1}. However, if  this symmetry would hold in discrete gravity, one could  hope for a unique anomaly--free (with repsect to diffeomorphisms) measure \cite{steinhaus}. As is also argued in \cite{steinhaus,dittinvariance}, the implementation of this symmetry into discrete gravity (of Regge type, i.e. with geometric data on a fixed triangulation or discretization), would make such a theory triangulation or discretization independent.  In this case there would also be no need of summing over triangulations, which is often employed to obtain a triangulation independent theory.

One can expect to find such a discretization independent theory for 3D gravity, which is a topological theory, i.e. there are no local physical (propagating) degrees of freedom. In fact we will succeed to find a triangulation invariant path integral description for 3D (linearized) Regge calculus. 4D gravity features local propagating degrees of freedom and a discretization independent model will require a non--local structure and moreover control over the solutions of the system \cite{song}. Nevertheless, as argued in \cite{steinhaus} the choice of path integral measure is important for the convergence of the model, also under a renormalization flow, which might be employed to find improved discretizations \cite{bahrdittrich2}. Moreover 4D classical Regge gravity is invariant under a set of certain local changes of the triangulation. One might therefore ask also for invariance of the path integral under this set of local changes.

In this work we will concentrate on finding a measure in a (Euclidean) Regge calculus set up, that is as much triangulation independent, as possible. Before explaining this in more detail we will shortly review  different measures suggested so far in the literature \cite{hambermeasure}. One method would be to discretize the (formal) continuum path integral
\ba\label{intro1}
I_{cont}=\int   \prod_{x, \,\rho \ge \tau} dg_{\rho\tau}(x) \,  \prod_x \left( \sqrt{\text{det}(g_{\mu\nu}})\right)^\alpha \,\, \exp\left( -S_{EH}\right) \q .
\ea
Here $S_{EH}$ is the (Euclidean) Einstein Hilbert continuum action and $ \left( \sqrt{\text{det}(g_{\mu\nu})}\right)^\alpha$ is a factor which can be obtained from the DeWitt metric on (geometric) superspace \cite{dewitt}.  More specifically the DeWitt measure \cite{dewitt} prescribes $\alpha=0$ in 4D and $\alpha=-1$ in 3D. However also other values of $\alpha$ have been suggested \cite{hambermeasure}, for instance $\alpha=-(D+1)$ for the Misner measure \cite{misner}, where $D$ is the dimension of space time. A priori it is not clear which choice to prefer \cite{hambermeasure}. 

Regge calculus \cite{regge} provides a discretization $S_R$ of the Einstein Hilbert action $S_{EH}$, defined on a triangulation. The metric data are replaced by edge lengths $l_e$ associated to the edges of the triangulation. As   $\sqrt{\text{det}(g_{\mu\nu})}$ gives the local space time volume a natural discretization of this factor is given by the volumes $V_\Delta$ of the top--dimensional simplices $\Delta$, i.e. 4--simplices in 4D and tetrahedra in 3D.  A straightforward discretization of (\ref{intro1}) is then given by \cite{hambermeasure} (modulo numerical constants)
\ba\label{intro2}
I_{discr} = \int  \prod_e dl_e^2 \prod_\Delta V^\alpha_\Delta \,\, \exp\left(-S_R\right)  \q .
\ea
Concerning the range of integration it will always be understood that the generalized triangle inequalities are satisfied. These require positive volume for all (sub-) simplices and are therefore equivalent to restricting the integration in the continuum path integral (\ref{intro1}) to positive definite metrics. 
Apart from this requirement of triangle inequalities (which are technically very difficult to implement) the measure used in(\ref{intro2}) has the advantage of being especially simple, in particular local.\footnote{Another suggestion is to use a measure of the form $\prod_e l_e^{-1}dl_e$, which is scale invariant. (The Regge action without cosmological constant term is invariant -up to an overall factor- under global rescaling of the edge lengths.) However, this measure did not lead to satisfying results in numerical simulations, see  \cite{lollreview} and references therein.} The simplicity is also a reason why $\alpha=0$ in 4D seems to be preferred \cite{hambermeasure}. 

In this paper we will consider a requirement of triangulation independence for the path integral measure. This requirement is also connected \cite{steinhaus} with a discrete notion of diffeomorphism invariance \cite{dittinvariance}. Hence asking for triangulation independence amounts to requiring an anomaly free measure with respect to the diffeomorphisms, see also \cite{martin_alex} for a discussion in the spin foam context.

Specifically we ask for invariance of the (linearized) model defined by (\ref{intro2}) under Pachner moves \cite{pachner}. These are local changes of the triangulation, that act ergodically, i.\ e.\ two topologically equivalent triangulations can always be transformed into each other by a sequence of Pachner moves. Restricting the measure to the local ansatz (\ref{intro2}) we will find that our results suggest to fix the parameter $\alpha$ to $\alpha=-\tfrac{1}{2}$ both in 3D and in 4D. Interestingly this conforms completely with the semi--classical analysis \cite{roberts, hellmannpr} of the Ponzano--Regge model \cite{pr} in 3D. This is a triangulation independent (spin foam) model for 3D quantum gravity, based on discrete variables. The case of 4D is much more involved. Firstly, being an interacting theory with propagating degrees of freedom, one cannot expect to obtain a triangulation independent model, with just local interactions, as in the Regge action \cite{song}. Indeed, we will precisely show in which sense the (linearized) 4D Regge action fails to be triangulation independent. Although  the semiclassical analysis of the 4D models \cite{nottingham,conrady} could show that the Regge action appears in a $\hbar \rightarrow 0$ limit of the amplitudes, the corresponding measure factor has not been specified yet as a function of the geometric variables.  For future work it will be interesting to compare in more detail the spin foam results with Regge gravity. Also a measure ambiguity shows up in choosing so called edge and face amplitudes \cite{martin_alex,ben_knotting,marseille}. These ambiguities could also be restricted by asking for as much triangulation independence as possible, similar to the method proposed here. Hence it would be very interesting to study the behavior of spin foam amplitudes under Pachner moves \cite{etera_valentin}.

There are also other suggestions for the Regge measure, which are non--local. As these are far more complicated explicit computations they have mostly been restricted to 2D. The Regge--Lund measure \cite{lund,hambergauge,hambermeasure} is obtained by discretizing first the deWitt super metric and then taking the determinant (whereas in (\ref{intro2}) this is performed the other way around). The result is given by
\ba
I_{RL} = \int \prod_e dl_e^2 \prod_\Delta  \sqrt{\text{det}(G_{ee'})} \,\, \exp\left(-S_R\right)  
\ea
where
\ba
G_{ee'}=-D! \sum_\Delta \frac{1}{V^\beta_\Delta} \,\frac{\partial V_\Delta}{\partial l_e l_{e'}}
\ea
and $\beta$ is another ambiguous parameter. As the determinant has to be taken of a matrix, which is indexed by all the edges of the triangulation, the result is potentially quite non--local. Further discussion of this measure can be found in \cite{hambergauge,hambermeasure}. 

In 3D, where gravity is a topological theory, we will find that a local measure is sufficient to guarantee triangulation independence of the (linearized) theory. In 4D, as previously mentioned one cannot expect to find complete triangulation independence for the path integral as already the action is not triangulation independent. (More precisely it is the Hamilton--Jacobi function as a functional of the boundary data, that is not invariant under the change of the bulk triangulation.) One can however ask for invariance under a restricted set of Pachner moves, under which the action happens to be invariant. These are the $4-2$ and $5-1$ moves (but not the $3-3$ move). Nevertheless also for these moves we will find that a factor appears that features a certain non-local structure. At this stage it seems however more promising to construct improved measures and actions directly by coarse graining and the method of perfect discretizations \cite{bahrdittrich2,song, steinhaus}. 

Ultimately, another criterion that any quantum gravity model has to satisfy, is to display the correct large scale limit. Also here a measure term could be essential. For investigations in 2D Regge see for instance \cite{ambjorn}, for discussion of the influence of the measure in the context of dynamical triangulations see \cite{scotts,dario-razvan}.  Another suggestion for constructing a measure for Regge gravity, is to mod out a certain subgroup of the continuum diffeomorphism group \cite{menotti}. This results again in a highly non--local measure, where explicit results are mostly restricted to 2D. 

\vspace{0.3cm}

In the next section we introduce the Regge action and its expansion up to second order. This requires the calculation of its Hessian matrix, which will be one of the main subjects of this work. Furthermore we discuss the concept of Pachner moves and briefly present the Pachner moves in 3D and 4D. Section \ref{sec:der_3d} deals with a general method to compute the Hessian matrix in 3D and presents the application of this method to the Regge actions associated to the  Pachner moves. Then we examine invariance of the path integral under Pachner moves and define a suitable measure factor. The results for 3D will be summarized in section \ref{sec:summary_3d}. In section \ref{sec:Hessian_4d} we extend our method to compute the Hessian matrix in 4D, examine invariance of the path integral under Pachner moves and discuss a suitable measure. The results in 4D are then summarized in section \ref{sec:summary_4d}. We conclude this work with a discussion of our results in section \ref{sec:discussion}.

\section{Linearized Regge Calculus} \label{sec:lin_regge}

The Regge action (which we will denote by $S$ in the following) provides a discretization of the Einstein Hilbert action for gravity. It is defined on a triangulation, the geometry is a piecewise flat one, and the geometric data are encoded in the lengths of the edges in this triangulation. For Regge type actions based on different geometric variables nearer to spin foams, specifically areas and angles, see \cite{area-angle, newregge}.

In the following we will consider the Euclidean path integral for the Regge discretization of gravity on a given 3D or 4D triangulation
\be
\int_{{l_e}_{|e\subset \partial M}}  \prod_{e\subset \text{bulk}} dl_e\, \mu(l_e) \exp\{-S\} \label{eq:amplitude} \q .
\ee
Here ${l_e}_{|e\subset \partial M}$ denotes the boundary conditions, which we take to be fixed length variables for the edges in the boundary triangulation. $\mu(l_e)$ is a suitable measure factor. In \eqref{eq:amplitude} not all edge lengths combinations are allowed since the edge lengths have to satisfy generalized triangle inequalities, i.e. all the (2D, 3D and, in case, 4D) volumes have to be positive. The (Euclidean) Regge action in arbitrary dimension $D$ can be written in the following form  
\be \label{eq:Regge_action}
S:=- \sum_{h \subset \text{bulk}} V_h    \left(2\pi  - \sum_{\sigma^D \supset h} \theta^{(\sigma^D)}_h\right)   - \sum_{h \subset \text{bdry}} V_h              \left(  \pi - \sum_{\sigma^D \supset h} \theta^{(\sigma^D)}_h \right)
%
%
%
\ee
where $\sigma^D$ denotes $D$-simplices, i.e. $D$-dimensional simplices with $D+1$ vertices, $h$ denotes `hinges', i.e. $D-2$-simplices, $V_h$ is the volume of a hinge and $\theta^{(\sigma^D)}_h$ denotes the internal dihedral angle in the $D$-simplex $\sigma^D$ at the hinge $h$. The terms in brackets in (\ref{eq:Regge_action}) define the bulk and boundary deficit angles
\ba
\omega^{\text{(bulk)}}_h:= 2\pi - \sum_{\sigma^D \ni h} \theta^{(\sigma^D)}_h \\
\omega^{\text{(bdry)}}_h:= k\pi - \sum_{\sigma^D \ni h} \theta^{(\sigma^D)}_h
\ea
where $k$ depends on the number of pieces one is glueing together at this boundary. If there are only two pieces we have $k=1$.


The dihedral angles are complicated functions of the lengths variables, so that the integral in (\ref{eq:amplitude}) cannot be computed analytically. Additionally one has to take the generalized triangle inequalities for the range of integration into account. 

To circumvent this issue, we consider linearized Regge Calculus in which one chooses a classical background solution (for the edge lengths) $l^{(0)}_e$ satisfying the triangle inequalities and one quantizes, i.e. integrates over, the perturbations $\lambda_e$ around it.

Therefore consider a small perturbation around a background solution
\be
l_e = l_e^{(0)} + \lambda_e
\ee
and expand the Regge action up to second order in the perturbation variables $\lambda_e$:
\be
S=S^{(0)}\Big|_{l_e=l^{(0)}_e} + \frac{\partial S}{\partial l_e}\Big|_{l_e=l^{(0)}_e} \lambda_e + \frac{1}{2} \frac{\partial^2 S}{\partial l_e \partial l_{e'}}\Big|_{l_e=l^{(0)}_e} \lambda_e \lambda_{e'} \q .\label{eq:Regge_expanded}
\ee
The background edge lengths $l^{(0)}_e$ are defined as the solution to the Regge equations,
\begin{equation}
 \frac{\partial S}{\partial l_e}=-\sum_{h \supset e} \frac{\partial V_h}{\partial l_e} \omega_h \;=\;0
\end{equation}
 such that the first order term in \eqref{eq:Regge_expanded} vanishes for the bulk edges. More specifically we take the background solution to be (locally) flat, that is $\omega^{\text{(bulk)}}_h=0$. (This is exactly the equation of motion in 3D.) The second order term is defined by  the matrix of second derivatives, that is the Hessian. 
 
 In three dimension one obtains due to the Schl\"afli identity\footnote{The Schl\"afli identity $\sum_{h\subset  \Delta^{(D)}} V_h\,\, \delta\theta^{(\Delta^{(D)})}_h =0$ ensures that terms with second derivatives of the dihedral angles vanish.}
\be
\frac{\partial^2 S}{\partial l_e \partial l_{e'}} =- \frac{\partial \omega_e}{\partial l_{e'}} \q .
\ee

In four dimensions we obtain (using again the Schl\"afli identity):
\be
\frac{\partial^2 S}{\partial l_e \partial l_{e'}} =- \sum_{h} \frac{\partial A_h}{\partial l_{e'}} \frac{\partial \omega_h}{\partial l_{e}} - \sum_{h \subset \text{bulk}} \frac{\partial^2 A_h}{\partial l_e \partial l_{e'}} \omega^{(\text{bulk})}_{h} - \sum_{h \subset \text{bdry}} \frac{\partial^2 A_h}{\partial l_e \partial l_{e'}} \omega_h^{(\text{bdry})} \q .
\ee
The bulk deficit angles  $\omega^{\text{(bulk)}}_h$ vanish on a flat (background) solution.

For the evaluation of these Hessian matrices, we will need the first derivatives of the dihedral angles with respect to the length variables.
A formula 
valid for simplices of arbitrary dimension $D$ can be found in \cite{Dittrich:2007wm}:
\begin{align}
\frac{\partial \tilde{\theta}_{kl}}{\partial l_{hm}}=& \frac{1}{D^2} \frac{l_{hm}}{\sin(\tilde{\theta}_{kl})} \frac{V_h V_m}{V^2} \Big( \cos(\tilde{\theta}_{kh}) \cos(\tilde{\theta}_{ml}) + \cos(\tilde{\theta}_{km}) \cos(\tilde{\theta}_{hl}) + \nn \\ 
& +\cos(\tilde{\theta}_{kl}) \big( \cos(\tilde{\theta}_{kh}) \cos( \tilde{\theta}_{km}) + \cos(\tilde{\theta}_{lh}) \cos( \tilde{\theta}_{lm})\big) \Big) \q .\label{eq:dih_der}
\end{align}
In \eqref{eq:dih_der} $\tilde{\theta}_{kl}$ denotes the dihedral angle (in a $D$-simplex) between the two $D-1$-simplices formed without the vertices $k$ and $l$ respectively. $l_{hm}$ is the length of the edge between vertices $h$ and $m$. $V_h$ denotes the volume of the $(D-1)$ simplex formed without vertex $h$ in the $D$-simplex. $V$ is the volume of the respective $D$-simplex.

In case the dihedral angle $\tilde{\theta}_{kl}$ and the edge $l_{hm}$ do not share a vertex, which implies that $(kl)=(hm)$, i.e. the hinge is formed without the vertices $h$ and $m$ in the $D$-simplex, equation \eqref{eq:dih_der} simplifies using the convention $\cos \tilde{\theta}_{ll} = -1$:
\begin{align}
\frac{\partial \tilde{\theta}_{hm}}{\partial l_{hm}} =& \frac{1}{D^2} \frac{l_{hm}}{\sin \tilde{\theta}_{hm}} \frac{V_h V_m}{V^2} \Big( 1 - \cos^2 \tilde{\theta}_{hm} \Big) \nn \\
=& \frac{1}{D^2} \frac{l_{hm} V_h V_m}{V^2} \sin \tilde{\theta}_{hm} = \frac{1}{D (D-1)} \frac{l_{hm} V_{hm}}{V} \label{eq:der_opposite} \q .
\end{align}
This result \eqref{eq:der_opposite}  will be crucial for an alternative 
 derivation of the matrix elements of the Hessian \eqref{eq:der_opposite} in section \ref{sec:der_3d}. This alternative derivation applies to configurations defining Pachner moves, which we will discuss in the next section.


\subsection{Pachner Moves} \label{sec:Pachner}

Pachner moves are local changes of the triangulation which, if applied consecutively,  allow to go from any triangulation of a given manifold to any other triangulation of that manifold \cite{pachner}. In quantum Regge calculus one usually fixes the triangulation and just integrates over the edge lengths in this given triangulation.  Given this definition the question arises of how the result depends on the choice of triangulation. Note that the triangulation is only an auxiliary structure, which is put in in order to regularize the (continuum) path integral. Hence it would be advantageous, if the path integral (with or without given boundary triangulation and condition) would depend minimally on the choice of (bulk) triangulation. In case the path integral does not depend at all on the triangulation, we do not even need to take any refinement limit (here of the bulk triangulation only), as the result will not change under refinement. Such a strong version of discretization independence can actually be expected in 3D, in which gravity is a topological theory, describing the dynamics of only global (topological) variables. Indeed we will find a measure that will render the path integral discretization independent in this sense. (That the linearized action is invariant under refinements has been shown in \cite{song}.)  Locally this discretization independence implies that the path integral is form invariant `under Pachner moves'. More precisely, we will consider here Pachner moves arising by integrating out certain edges in the triangulation, so that the remaining edges still define a triangulation. The form of the (discretized) path integral should then be invariant.

We will consider a similar requirement in 4D. This defines however, a theory with local degrees of freedom, where not even the (linearized) action is invariant under change of triangulation \cite{bahrprivate,song,philipp}. This broken invariance can however be isolated into one of the Pachner moves, the $3-3$ move. Hence we can at least ask whether it is possible to define a measure that would render the path integral invariant under the remaining Pachner moves.


A $x-y$ Pachner move changes a complex of $x$ $D$--simplices into one with $y$ $D$--simplices without changing the boundary triangulation. Here the parameters $x,y$ are related by $x+y=D+2$. Since the boundary is not changed, the Pachner moves act locally in the triangulation.  This also allows us (for the cases with $x>y$) to consider the initial configuration of $x$ $D$ simplices, to integrate out the bulk edges and to re--interpret the resulting partition function as one for the complex with $y$ simplices. Note also that the $x-y$ and the $y-x$ move are inverse to each other.

In the following sections we will introduce the Pachner moves in 3D and 4D and shortly point out some points pertaining to the dynamics defined by Regge calculus.  

\subsection{Pachner moves in 3D} \label{sec:Pachner_3d}

Here we have two Pachner moves $3-2$ and $4-1$ (and their inverses). Note that the equation of motion for 3D Regge calculus require flatness, i.e. vanishing deficit angles.

\subsubsection{$3-2$ move} \label{sec:Pachner_3-2}

\begin{figure}[h]
\centering
\includegraphics[scale=0.7]{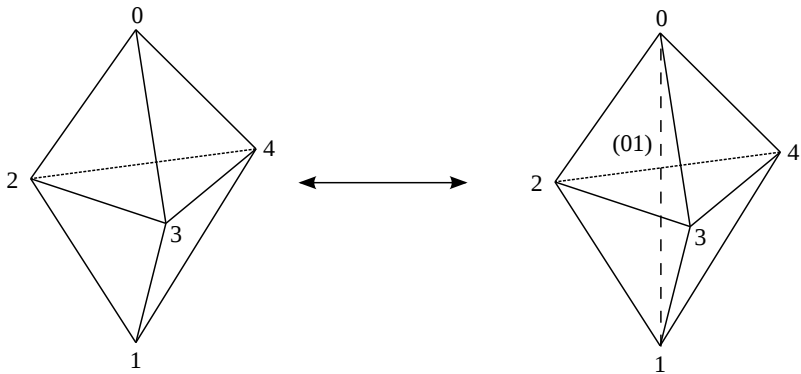}
\caption{$3 - 2$ move. The two tetrahedra can be split into three by connecting the two vertices separated by the shared triangle. The dashed line in the three tetrahedra configuration is the dynamical edge.}
\label{fig:3-2}
\end{figure}

The first Pachner move we will  consider  is the $3- 2$ move, see Fig. \ref{fig:3-2}.  In the initial configuration three tetrahedra $(0 1 2 3)$, $(0 1 2 4)$ and $(0 1 3 4)$ share an edge $(01)$. This is the only bulk edge. Removing (i.e. integrating out) this edge and introducing a triangle $(1 2 3)$ we obtain a configuration of two tetrahedra $(0 2 3 4)$ and $(1 2 3 4)$ sharing this triangle.

As there is only one bulk edge in the initial configuration, we will also have only one equation of motion. This equation of motion requires the vanishing of the bulk deficit angle $\omega_{01}$ and in this way fixes the length $l_{01}$ of the edge $(01)$ as a function of the boundary edge lengths. 


\subsubsection{$4-1$ move} \label{sec:Pachner_4-1}

\begin{figure}[h]
\centering
\includegraphics[scale=0.7]{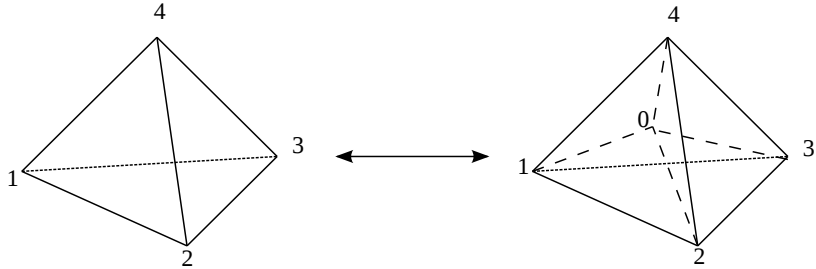}
\caption{$4 - 1$ move. The tetrahedron is split into four by placing one additional vertex inside the tetrahedron and connecting it to the remaining vertices in the boundary giving four internal edges (dashed).}
\label{fig:4-1}
\end{figure}

The other Pachner move in 3D we are going to discuss is the $4-1$ move. Here in the initial configuration four tetrahedra share one vertex. This configuration can be obtained by subdividing a tetrahedron $(1 2 3 4)$ into four tetrahedra by placing one vertex $0$ into the tetrahedron and connecting $0$ with the other four vertices. In the $4-1$ move this vertex $0$ and the adjacent edges are removed, leaving us with one tetrahedron $(1 2 3 4)$, see also Fig. \ref{fig:4-1}. 

In the initial configuration with four tetrahedra, there are four bulk edges, and hence four equations of motion. These, again require that the (four) bulk deficit angles have to vanish, i.e. that the complex has to be flat. We know that we can easily construct such solutions by placing a vertex into the (flat) tetrahedron $(1 2 3 4)$ and determining the lengths of the four bulk edges. There is, of course, a three--dimensional parameter space of where to place the inner vertex exactly, hence the solutions are not uniquely determined. This is the well known gauge freedom in Regge calculus on flat solutions \cite{roceck, morse, Freidel:2002dw, Dittrich:2007wm,dittrichreview}, a discrete remnant of the diffeomorphism symmetry in the continuum. From this it follows that of the four equations of motions only one is independent and that we have to expect three null modes in the Hessian matrix of the system, signifying three gauge degrees of freedom. Further discussions and extensions to the case with cosmological constant can be found in \cite{dittrichreview,bahrdittrich1,bahrdittrich2}.


\subsection{Pachner moves in 4D} \label{sec:Pachner_4d}

\subsubsection{$4- 2$ move} \label{sec:Pachner_4-2}

\begin{figure}[h]
\centering
\includegraphics[scale=0.7]{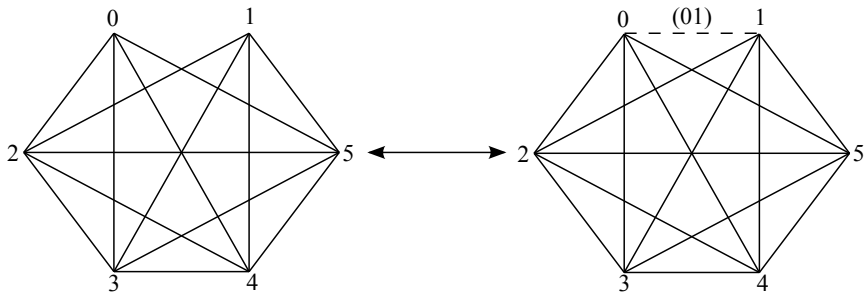}
\caption{$4 - 2$ move. By connecting the vertices $(0)$ and $(1)$ the two 4-simplices are split into four with one bulk edge, here drawn dashed.}
\label{fig:4-2}
\end{figure}

This Pachner move is very similar to the $3- 2$ move in 3D, see Fig. \ref{fig:4-2}. 
The initial configuration is one with four 4--simplices, $(0 1 2 3 4)$, $(0 1 2 3 5)$, $(0 1 2 4 5)$ and $(0 1 3 4 5)$, sharing one edge $(01)$. All the other edges are boundary edges. Removing this edge and introducing a tetrahedron $(2 3 4 5)$, we obtain a configuration with two $4$-simplices $(0 2 3 4 5)$ and $(1 2 3 4 5)$ sharing this tetrahedron. 

As there is only one bulk edge we again have only one equation of motion for the initial configuration. A (flat) solution can always be constructed in the following way: The boundary triangulation is the same as for two 4--simplices sharing one tetrahedron. Such a configuration can always (i.e. for all boundary edge lengths satisfying the appropriate inequalities) be embedded into flat 4D space. We can hence straightforwardly determine the distance between the vertices $0$ and $1$ in the induced metric, which defines the length of the edge $(01)$. 

In some exceptional cases there might be also solutions with curvature \cite{bahrdittrich1}, however this seems to be rather a discretization artifact. For the perturbative solutions around flat space we are interested in, we can note that the linearized equations of motion have a unique (flat) solution for all (linearized) boundary perturbations.


\subsubsection{$5-1$ move} \label{sec:Pachner_5-1}

\begin{figure}[h]
\centering
\includegraphics[scale=0.7]{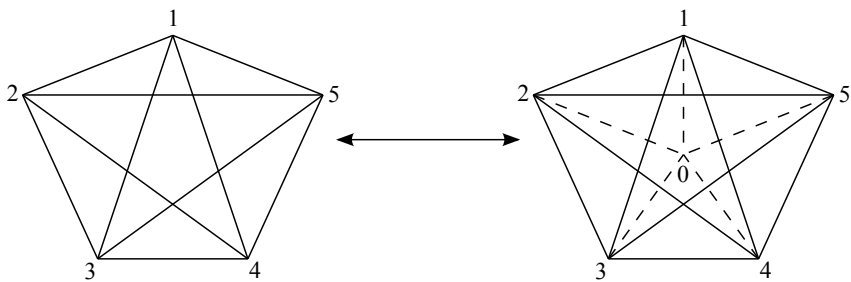}
\caption{$5 - 1$ move. The 4-simplex is split into five 4-simplices by placing one vertex inside the 4-simplex and connecting it to the boundary vertices, hence obtaining five bulk edges (dashed lines).}
\label{fig:5-1}
\end{figure}

The $5-1$ is again analogous to the $4-1$ move in 3D. In the initial configuration five 4--simplices share one vertex $(0)$ which is adjacent to five bulk edges. Removing this vertex and the adjacent edges we are left with just one simplex $(12345)$, see also Fig. \ref{fig:5-1}

Also here, we can construct for all boundary configurations flat solutions to the equations of motion. These can be found by placing the vertex $(0)$ into the (flat) 4--simplex  $(12345)$ and determining the induced lengths of the edges $(0x)$, where $x=1,\ldots,5$. For given boundary data there is a four--parameter space of such solutions, according to the  four parameters describing the position of the vertex inside the 4--simplex. Hence we can expect four null modes for the Hessian of this configuration.


\subsubsection{$3 - 3$ move} \label{sec:Pachner_3-3}

We are left with the $3-3$ move, which is significantly different from all the other Pachner moves discussed so far. 

Assume three 4-simplices $(0 1 2 3 4)$, $(0 1 2 3 5)$ and $(0 1 2 4 5)$ sharing one triangle $(0 1 2)$. Note that this configuration does not include a triangle $(3 4 5)$, as neither of the three 4--simplices contains the three vertices $(3),(4),(5)$.

\begin{figure}[h]
\centering
\includegraphics[scale=0.7]{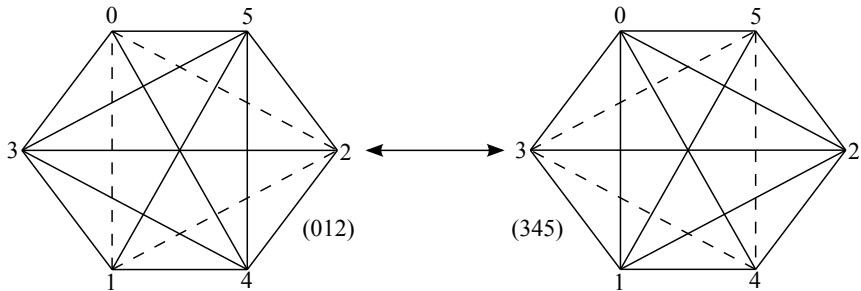}
\caption{$3 - 3$ move. Three 4--simplices sharing the triangle $(0 1 2)$ and not containing $(3 4 5)$ are rebuilt into three 4--simplices sharing the triangle $(3 4 5)$ and not including triangle $(0 1 2)$. The shared triangles are drawn dashed in this figure. Note that all edges are boundary edges and are contained in both configurations, so the configurations are determined by the shared triangle.}
\label{fig:3-3}
\end{figure}

The $3-3$ move rebuilds this configuration into three 4--simplices $(0 1 3 4 5)$, $(0 2 3 4 5)$ and $(1 2 3 4 5)$ which share the triangle $(3 4 5)$ and do not include the triangle $(0 1 2)$, see also Fig. \ref{fig:3-3}.

In contrast to all other Pachner moves discussed so far the $3- 3$ move does not involve dynamical edges, i.e. all edges are in the boundary and therefore included in both configurations. We therefore do not have an equation of motion. Note however that, again in contrast to the other Pachner moves, not all boundary configurations define a flat geometry. That is, in both configurations we have only one bulk triangle. The vanishing of the deficit angle for this bulk triangle gives one condition for the length of the boundary edges. In case this condition is violated we do have a curved configuration. In particular, even on a flat background, we can have a curvature excitation, if the boundary perturbations do not satisfy the linearized flatness condition. 


\vspace{0.2cm}

\noindent 

In the following section we will specify the Hessian matrix of the Regge action associated to the various configurations appearing in the Pachner moves. We could start with the formula (\ref{eq:dih_der}) for the derivatives of the dihedral angles to obtain the derivatives of the deficit angles, so that these can be combined to give the entries in the Hessian. This procedure would however result in very lengthy formulas and not use the flatness of the background solution. We will use an alternative strategy, which will produce a quite enlightening structure for the Hessian, and for which we present some auxiliary formulas in the next section.

\section{Computation of the Hessian matrix in 3D} \label{sec:der_3d}

In this section we will compute the matrix elements of the Hessian matrix. To do so one has to compute terms of the form $\frac{\partial \omega}{\partial l}$ for which we will present a general strategy, similar to \cite{Korepanov:2000jp,Korepanov:2000aj}. We need to extend the ideas in  \cite{Korepanov:2000jp,Korepanov:2000aj} in order to also obtain the matrix elements of the Hessian indexed by edges in the boundary. First we will derive two auxiliary formulas, which is the subject of the next section.

\subsection{Auxiliary formulas} \label{sec:aux_form_3d}

For concreteness we will derive the auxiliary formulas for the initial configuration of the $3 - 2$ move (see Fig. \ref{fig:3-2}) with the bulk edge $(0 1)$ as described in the previous section. Assume that this configuration can be embedded into flat (3-dimensional) space, i.e. $\mathbb{R}^3$. This implies that for instance $l_{01}$, the edge length of the dynamical edge $(01)$, is fixed as a function of all other edge lengths. Hence there is one relation which all edge lenghts have to satisfy, which is $\omega_{01}=0$, i.e. the deficit angle at the edge $(01)$ vanishes. (Note that this relation can also be derived by requiring that the Cayley-Menger determinant of this configuration, giving the square of the 4D volume, vanishes.) As there is one condition, at least two edge lenghts have to be varied in order to preserve this relation. Therefore consider variations of exactly two edge lenghts, $l$ and $l'$. Alternatively one can interpret $l'$ as a function of $l$, where all other edge lengths are (fixed) parameters. We will derive a relation between the edge variations $\delta l$ and $\delta l'$ or alternatively the partial derivative $\frac{\partial l'}{\partial l}$.

For example, consider $l'=l_{23}$ and $l=l_{34}$. We vary $l_{23}$ and $l_{34}$ such that $\omega_{01}=0$, i.e. the triangulation is still embeddable in flat space. This implies:
\begin{align}
0 = \delta \omega_{01} = & - \delta \theta_{01}^{(0123)} - \delta \theta_{01}^{(0124)} - \delta \theta_{01}^{(0134)} \nn \\
= & - \delta \theta_{01}^{(0123)} - \delta \theta_{01}^{(0134)} \label{eq:der_edge_rel1}
\end{align}
where $\theta_{01}^{(01xy)}$ is the dihedral angle at edge $(01)$ in the tetrahedron $(0 1 x y)$. $\delta \theta_{01}^{(0 1 2 4)}=0$ since it neither depends on $l_{23}$ nor on $l_{34}$,  as these edges are not part of the tetrahedron $(0124)$.   


\noindent Using equation \eqref{eq:der_opposite}, we obtain:
\be
\frac{\partial \theta^{(0123)}_{01}}{\partial l_{23}}  \,=\, \frac{l_{01} l_{23}}{6 V_{\bar{4}}} \label{eq:opp_ang_1} \q, \q\q
\frac{\partial \theta^{(0134)}_{01}}{\partial l_{34}}  \,=\, \frac{l_{01} l_{34}}{6 V_{\bar{2}}}            
\ee
where $V_{\bar{i}}$ denotes the volume of the tetrahedron formed by all vertices except $i$, e.g. $\bar{4} \rightarrow (0 1 2 3)$, such that $V_{\bar{4}} = V_{(0 1 2 3)}$. 
Since $\delta \theta_{01}^{(0 1 2 3)}$  ($\delta \theta_{01}^{(0 1 3 4)}$) can only depend on $l_{23}$  ($l_{34}$ respectively), we can use equations \eqref{eq:opp_ang_1} in \eqref{eq:der_edge_rel1} and obtain:
\begin{equation}
\frac{\partial l_{23}}{\partial l_{34}} = - \frac{l_{34}}{l_{23}} \frac{V_{\bar{4}}}{V_{\bar{2}}}  \q .
\end{equation}
In general one finds (for a five vertex configuration with vanishing Cayley-Menger determinant) \cite{Korepanov:2000jp,Korepanov:2000aj}:
\begin{equation} \label{eq:edge_rel1}
\left|\frac{\partial l_{ij}}{\partial l_{jk}}\right| = \left|\frac{l_{jk} V_{\bar{k}}}{l_{ij} V_{\bar{i}}}\right|  \q .
\end{equation}
The actual sign depends on the geometric configuration under consideration.

In addition to relation \eqref{eq:edge_rel1} we need an analogous relation between deviations of edges not sharing a vertex. This can be derived from \eqref{eq:edge_rel1}: Consider variations of three edge lengths $l_{ij}$, $l_{jk}$ and $l_{km}$ such that $\omega_{01}=0$. That is, $l_{ij}$ can be understood as a function of $l_{jk}$ and $l_{km}$. Then
\begin{equation} \label{eq:no_v_shared1}
\delta l_{ij} = \frac{\partial l_{ij}}{\partial l_{jk}} \delta l_{jk} + \frac{\partial l_{ij}}{\partial l_{km}} \delta l_{km}  \q .
\end{equation}
Now we restrict the variations further by requiring $\delta l_{ij}=0$, such that we have to additionally understand $l_{jk}$ as a function of $l_{km}$, that is $l_{ij}=l_{ij}\Big(l_{jk}(l_{km}),l_{km}\Big)$. Thus one obtains for \eqref{eq:no_v_shared1}:
\begin{align}
0 =& \frac{\partial l_{ij}}{\partial l_{jk}} \frac{\partial l_{jk}}{\partial l_{km}} \delta l_{km} + \frac{\partial l_{ij}}{\partial l_{km}} \delta l_{km} \\
\implies & \frac{\partial l_{ij}}{\partial l_{km}} =  - \frac{l_{ij}}{l_{jk}} \frac{l_{jk}}{l_{km}} \q .
\end{align}
With \eqref{eq:edge_rel1} we find (see also \cite{Korepanov:2000aj}):
\begin{equation} \label{eq:edge_rel2}
\left|\frac{\partial l_{ij}}{\partial l_{km}} \right| = \left| \frac{l_{km} V_{\bar{k}} V_{\bar{m}}}{l_{ij} V_{\bar{i}} V_{\bar{j}}}\right|  \q .
\end{equation}

To summarize \eqref{eq:edge_rel1}, \eqref{eq:edge_rel2}:
\begin{equation*}
\left| \frac{\partial l_{ij}}{\partial l_{jk}} \right| =\left| \frac{l_{jk} V_{\bar{k}}}{l_{ij} V_{\bar{i}}} \right|\;\;, \;\;\;\;
\left| \frac{\partial l_{ij}}{\partial l_{km}} \right| = \left| \frac{l_{km} V_{\bar{k}} V_{\bar{m}}}{l_{ij} V_{\bar{i}} V_{\bar{j}}}\right|
\end{equation*}
where in fact \eqref{eq:edge_rel1} is a special case of \eqref{eq:edge_rel2}.

In the following section we will use relations \eqref{eq:der_opposite}, \eqref{eq:edge_rel1} and \eqref{eq:edge_rel2} to compute terms of the form $\frac{\partial \omega}{\partial l}$.

\subsection{Computation of $\frac{\partial \omega}{\partial l}$} \label{sec:der_omega_3d}

The Hessian of the Regge action has entries of the form $\frac{\partial \omega}{\partial l}$, which we have to evaluate on configurations where $\omega^{(bulk)}=0$. 
As in the previous section we consider the initial configuration of the $3 - 2$ move.
We will start with the calculation of $\frac{\partial \omega_{01}}{\partial l_{01}}$, which is the derivative of the bulk deficit angle with respect to the bulk edge length.

The equation of motion for the perturbations $\lambda_{01}$ around the flat solution $l_{01}^{(0)}$ is given by 
\ba\label{bia1}
0\;=\;\sum_b \frac{\partial^2 S}{\partial l_b \partial l_{01}} \lambda_b + \frac{\partial^2 S}{\partial l_{01} \partial l_{01}} \lambda_{01}
=-\sum_b \frac{\partial \omega_{01}}{\partial l_{b}} \lambda_b -\frac{\partial \omega_{01}}{\partial l_{01}} \lambda_{01}  \q .
\ea
Here $b$ indicates edges in the boundary triangulation, and the sum is over all such edges.

Now we know that these equation of motion specify $\lambda_{01}$, such that the linearized deficit angle at $(01)$ is still flat. That is, if we choose the boundary perturbations, such that for instance only $\lambda_{23} \sim \delta l_{23}$ is non--vanishing, we know that the ration of $\lambda_{01}\sim  \delta l_{01}$ and $\lambda_{23} \sim \delta l_{23}$ has to satisfy  \eqref{eq:edge_rel2}. This specifies the ratio of the derivatives $\partial \omega_{01}/\partial l_{01}$ and $\partial \omega_{01} /\partial l_{23}$. For the latter derivative, as $l_{23}$ is only included in one of the three tetrahedra we have via (\ref{eq:der_opposite})
\be\label{bia2}
\frac{\partial \omega_{01}}{\partial l_{23}}\,=\,-\frac{\partial \theta^{(0123)}_{01}}{\partial l_{23}}\,=\, -\frac{l_{01} l_{23}}{6 V_{\bar{4}}}  \q .
\ee
%
This finally gives
\begin{equation} \label{eq:deficit_bulk_bulk}
\left|\frac{\partial \omega_{01}}{\partial l_{01}}\right| = \left|\frac{l_{01} l_{23}}{6 V_{\bar{4}}} \frac{\delta l_{23}}{\delta l_{01}} \right| \underset{\eqref{eq:edge_rel2}}{=} \left| \frac{l_{01}^2}{6} \frac{V_{\bar{0}} V_{\bar{1}}}{V_{\bar{2}} V_{\bar{3}} V_{\bar{4}}} \right| \q .
\end{equation}
The actual sign is determined by the geometry and discussed in the next section. 
 Note that we could have also used the lengths $l_{24}$ or $l_{34}$ instead of $l_{23}$, which would have however all lead to the same result.

Next we consider terms of the form $\frac{\partial \omega_{01}}{\partial l_{b}}$, i.e. derivatives of the deficit angle at the bulk edge with respect to a boundary edge length. Note that for $b=23,24,34$ the result is already given by (the analogue of) (\ref{bia2}). 

To find the derivative with respect to the remaining boundary lengths consider again (\ref{bia1}) with all boundary perturbations vanishing except, say $\lambda_{0i}\sim \delta l_{0i}$. Then, with the same line of arguments as used previously we can conclude 
%
\begin{equation} \label{eq:der_deficit_bdry}
 \frac{\partial \omega_{01}}{\partial l_{0i}} = - \frac{\partial \omega_{01}}{\partial l_{01}} \frac{\delta l_{01}}{\delta_{0i}}
\end{equation}
and hence 
\begin{equation} \label{eq:der_bulk_bound_sh}
\left| \frac{\partial \omega_{01}}{\partial l_{0i}} \right| \underset{\eqref{eq:edge_rel1}}{=} \left| \frac{l_{01} l_{0i}}{6} \frac{V_{\bar{0}}}{V_{\bar{j}} V_{\bar{k}}} \right|
\end{equation}
where $i\, \in \,\{2,3,4\}$ and
 $j,\,k $ are such that ${i,j,k}={2,3,4}$. Again, the sign is determined by the geometry under consideration.


Note that due to the symmetry of second derivatives of the Regge action 
we have
\begin{equation}
\frac{\partial \omega_e}{\partial l_{e'}} = \frac{\partial \omega_{e'}}{\partial l_e}\,.
\end{equation}
Hence we can deduce terms of the form $\frac{\partial \omega_{b}}{\partial l_{01}}$ from $\frac{\partial \omega_{01}}{\partial l_{b}}$. Thus only terms of the form $\frac{\partial \omega{b}}{\partial l_{b'}}$ remain to be computed, i.e. derivatives of exterior angles with respect to boundary edge lengths.

To this end, remember that the initial configuration of the $3-2$ Pachner move is flat. During the Pachner move the edge $(01)$ is removed and replaced by a triangle $(234)$, such that neither the intrinsic geometry (i.\ e.\ flatness) nor the extrinsic geometry (the embedding into flat space) of the boundary changes. In particular we will have that the extrinsic curvature angles $\omega_b^{(3)}(l_{b'},l_{01}(l_{b'}))=\omega_b^{(2)}$ coincide in the initial and finial configuration of the Pachner moves, involving three or two tetrahedra respectively. Here we understand $l_{01}$ as a function of the boundary lengths $l_{b'}$ as it is determined by the requirement of flatness.


Now varying just one boundary edge lengths $l_{b'}$, together with $l_{01}=l_{01}(l_{b'})$ as function of this lengths we obtain:
\begin{align}
&\frac{d \omega^{(2)}_b}{d l_{b'}} = \frac{d \omega^{(3)}_{b}}{d l_{b'}}  = \frac{\partial \omega^{(3)}_{b}}{\partial l_{b'}} + \frac{\partial \omega^{(3)}_{b}}{\partial l_{01}} \frac{\partial l_{01}}{\partial l_{b'}} \\
\implies & \frac{\partial \omega^{(3)}_{b}}{\partial l_{b'}} = \underbrace{\frac{d \omega^{(2)}_{b}}{d  l_{b'}}}_{\frac{\partial \omega^{(2)}_{b}}{\partial l_{b'}}} - \underbrace{\frac{\partial \omega^{(3)}_{b}}{\partial l_{01}}}_{\frac{\partial \omega_{01}}{\partial l_{b'}}} \frac{\partial l_{01}}{\partial l_{b'}} \q .
\end{align}
This gives finally
\ba
\frac{\partial \omega^{(3)}_b}{\partial l_{b'}} = \frac{\partial \omega^{(2)}_b}{\partial l_{b'}} + s \, \frac{l_{b} l_{b'}}{6} \frac{V_{\bar{i}} V_{\bar{j}} V_{\bar{m}} V_{\bar{n}} }{\prod_n V_{\bar{n}}}
\ea
for $b=(ij)$ and $b'=(mn)$. Here $s=\pm 1$ denotes a sign, that will be determined in the next section.

\subsubsection{Determining the sign of $\frac{\partial l_e}{\partial l_{e'}}$} \label{sec:sign_3d}

In the previous section we have seen that in order to compute the full expression for the matrix elements of the Hessian, the actual sign of the derivatives of the form $\frac{\partial l_e}{\partial l_{e'}}$ has to be determined. To be more precise, one only needs to determine the signs of $\frac{\partial l_{01}}{\partial l_{b}}$, where one has to treat the cases in which $l_{b}$ shares a vertex with $l_{01}$ and where it does not share a vertex separately.

We start with the case where $l_{b}$ shares a vertex with $l_{01}$, e.g. $l_{0i}$ with $i \in \{2,3,4\}$. In the derivation of the formula for $\frac{\partial \omega_{01}}{\partial l_{0i}}$, we considered variations of the edge lengths $l_{01}$ and $l_{0i}$, while keeping all other edge lengths fixed, under the condition that the triangulation is supposed to remain flat. This allowed us to understand $l_{01}$ as a function of $l_{0i}$, $l_{01} = l_{01}(l_{0i})$. To determine the sign of this dependence, consider Fig. \ref{fig:edge_var_bulk}:

\begin{figure}[h]
\centering
\includegraphics[scale=0.7]{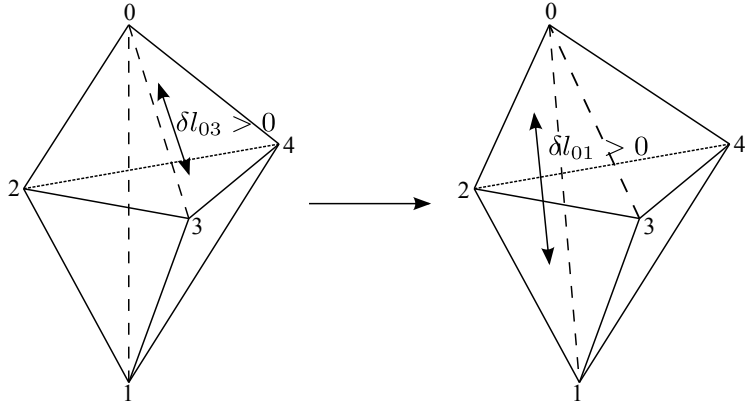}
\caption{As one increases the edge lengths $l_{03}$, one also has to increase $l_{01}$ in order to keep the triangulation flat, i.e. $\omega_{01} = 0$} \label{fig:edge_var_bulk}
\end{figure}

Assume that we enlarge $l_{0i}$ slightly, i.e. $\delta l_{0i} >0$. If we do not change $l_{01}$ as well, the condition $\omega_{01} = 0$ will be violated since all other edge lengths are fixed. However, if one allows $l_{01}$ to vary as well, the vertex $(0)$ will be `pushed' away from the vertex $i$, but since the edge lengths $l_{0j}$ and $l_{0k}$ are fixed, $l_{01}$ has to be increased, i.e. $\delta l_{01} >0$. Hence:
\begin{equation}
\delta l_{0i} > 0 \, \implies \, \delta l_{01} >0 \underset{\eqref{eq:edge_rel1}}{\implies} \frac{\delta l_{01}}{\delta l_{0i}} = \frac{l_{0i} V_{\bar{i}}}{l_{01} V_{\bar{1}}} \q .
\end{equation}

We follow a similar line of argumentation for terms of the form $\frac{\partial l_{ij}}{\partial l_{01}}$, with $i,j \, \in \, \{2,3,4\}$. Consider Fig. \ref{fig:edge_var_bdry}:

\begin{figure}[h]
\centering
\includegraphics[scale=0.7]{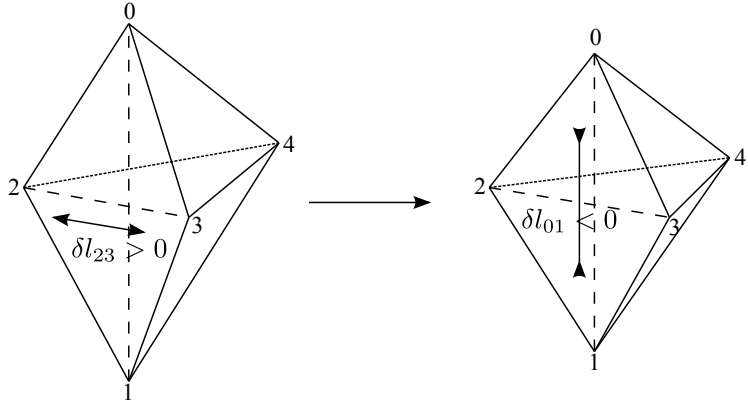}
\caption{As one increases the edge lengths $l_{23}$, one has to decrease $l_{01}$ in order to keep the triangulation flat, i.e. $\omega_{01}=0$} \label{fig:edge_var_bdry}
\end{figure}

Assume that we slightly increase $l_{ij}$, i.e. $\delta l_{ij} > 0 $. Since the edge lenghts $l_{ik}$ and $l_{jk}$ are fixed, the vertex $k$ is being `pulled' towards the edge $(ij)$. Furthermore the edge lengths $l_{0i}$, $l_{0j}$ and $l_{0k}$ are fixed, such that the vertex $(0)$ is `dragged' towards the edge $(ij)$. This configuration can only remain flat, if $l_{01}$ is decreased, i.e. $\delta l_{01} < 0$. Hence:
\begin{equation}
\delta l_{ij} >0 \, \implies \, \delta l_{01} <0 \, \underset{\eqref{eq:edge_rel2}}{\implies} \, \frac{\delta l_{01}}{\delta l_{ij}} = - \frac{l_{ij} V_{\bar{i}} V_{\bar{j}}}{l_{01} V_{\bar{0}} V_{\bar{1}}} \q .
\end{equation}
This is also consistent with \eqref{bia2}.

\subsection{Summary for the $3 - 2$ move} \label{sec:sum_3-2}

Let us summarize the results of the previous paragraphs.
\begin{itemize}
\item In case either edge $(ij)$ or the edge $(km)$ is in the bulk, one obtains:
 \begin{equation} \label{eq:Hess_bulk}
   H^{(3)}_{(ij),(km)} := \frac{\partial^2 S}{\partial l_{km} \partial l_{ij}} =- \frac{\partial \omega_{ij}}{\partial l_{km}} = (-1)^{s_{i} + s_j + s_k + s_{m}} \frac{l_{ij} l_{km}}{6} \frac{V_{\bar{i}} V_{\bar{j}} V_{\bar{k}} V_{\bar{m}}}{\prod_{n} V_{\bar{n}}}
  \end{equation}
  where 
  \begin{equation}
   s_i = \left\{
   \begin{array}{l r}
   1 & \quad \text{if }i \in \{0,1\} \\
   0 & \quad \text{else}
   \end{array}
   \right.
  \end{equation}
  and the product in the denominator runs over all vertices in the triangulation.
\item In case both the edges $(ij)$ and $(km)$ are in the boundary, one obtains:
 \begin{equation} \label{eq:Hess_bdry}
   H^{(3)}_{(ij),(km)} := \frac{\partial^2 S}{\partial l_{km} \partial l_{ij}} = - \frac{\partial \omega^{(3)}_{ij}}{\partial l_{km}} =(-1)^{s_{i} + s_j + s_{k} + s_m} \frac{l_{ij} l_{km}}{6} \frac{V_{\bar{i}} V_{\bar{j}} V_{\bar{k}} V_{\bar{m}}}{\prod_{n} V_{\bar{n}}} - \frac{\omega^{(2)}_{ij}}{\partial l_{km}}
  \end{equation}
 where $\omega^{(i)}_{km}$ denotes the exterior angle at the (boundary) edge $(km)$ in the $i$ tetrahedra configuration, $s_i$ is defined as above.
\end{itemize}

Notice the simple form of the Hessian,
\ba\label{bia210}
H^{(3)}_{(ij),(km)} = H^{(2)}_{(ij),(km)} + c \,h_{(ij)} \,h_{(km)} \q ,
\ea
in particular that the second summand in (\ref{bia210}) factorizes. (Here $H^{(2)}_{(ij),(km)} =0 $ if either $(ij)$ of $(km)$ equals $(01)$.)

We have now all the prerequisites  to discuss the (form-) invariance of the path integral associated to the $3-2$ move.

\subsection{Invariance of the path integral} \label{sec:inv_pi_3-2}

For the $3 - 2$ move we have to consider an expression of the following form:
\begin{equation} \label{eq:3-2_pi}
P_{3 - 2} = \int d \lambda_{01} \; \mu(l) \; \exp\left[- \sum_{(ij),(km)} \frac{1}{2} H^{(3)}_{(ij),(km)}\, \lambda_{ij} \, \lambda_{km} \right]
\end{equation}
where:
\begin{itemize}
 \item $\mu(l)$ is a measure factor, which we assume to only depend on the background variables $l$, such that the configuration is flat.
 \item $H^{(3)}_{(ij),(km)}$ is the $(ij)$$(km)$-matrix element of the Hessian in the three tetrahedra configuration, which we computed in the previous section, see \eqref{eq:Hess_bulk}, \eqref{eq:Hess_bdry}.
 \item Since $\lambda_{01}$ is the only dynamical edge (variation) in the configuration under discussion, the sign of $H^{(3)}_{(01),(01)}$ is crucial for the convergence of \eqref{eq:3-2_pi}. In the sign convention introduces in equation \eqref{eq:Regge_action} $H^{(3)}_{(01),(01)} >0$, such that \eqref{eq:3-2_pi} converges.
\end{itemize}

\noindent  We can easily perform the integral in \eqref{eq:3-2_pi} as it is a (partial) Gaussian integration. 
 For an integral of the form 
\begin{equation}
I = \int dq_1 \, \ldots \, dq_r \; \exp\left\{ -\frac{1}{2} \vec{q}^T \, M \, \vec{q} \right\}
\end{equation}
where $M$ is a real, symmetric, positive-definite $n \times n$-matrix and $\vec{q}=(q_i)$ denotes a vector with $i=1,\ldots,r,r+1,\ldots n$. 
Splitting the  matrix $M$ accordingly  into submatrices
\begin{equation}
M=
\left( 
\begin{array}{c c}
\; W_0 \; &\; V \; \\
\; V^T \; &\; U_0 \;
\end{array}
\right)
\end{equation}
we can write
\begin{equation}
I = \frac{(2 \pi)^{\frac{r}{2}}}{\sqrt{\det(W_0)}} \, \exp\left\{ - \frac{1}{2}\, \vec{u}^T\, U \, \vec{u} \right\}
\q\q\text{with}\q\q
U := U_0 - V^T \, W_0^{-1} \, V \q .
\end{equation}
Using this result for the $3-2$ move  (\ref{eq:3-2_pi}), we identify:
\begin{align}
W_0 \;=\; H^{(3)}_{(01),(01)} \; = \; &- \frac{\partial \omega_{01}}{\partial l_{01}} \\
(U_0)_{b,b'} \;=\; &\underbrace{-\frac{\partial \omega^{(2)}_{b}}{\partial l_{b'}}}_{H^{(2)}_{b,b'}} + \frac{\partial \omega_{01}}{\partial l_{b}}\frac{\partial l_{01}}{\partial l_{b'}} \\
(V)_{(01),b} \; = \; & - \frac{\partial \omega_{01}}{\partial l_{b}}   \q .
\end{align}
Furthermore requiring form invariance of \eqref{eq:3-2_pi} implies
\begin{equation}
P_{3 - 2} \propto \exp \left\{ - \sum_{(ij)\neq (01),(km) \neq (01)}\frac{1}{2} \, H^{(2)}_{(ij)(km)}\, \lambda_{ij} \, \lambda_{km} \right \}
\end{equation}
such that in order to show that \eqref{eq:3-2_pi} is invariant (on the level of the action), one has to show that
$U \, = \, H^{(2)}$,
which implies
\begin{equation} \label{eq:form_inv_1}
- \frac{\partial \omega_{01}}{\partial l_{b}} \frac{\partial l_{01}}{\partial l_{b'}}\, = \, \left[ \frac{\partial \omega_{01}}{\partial l_{01}} \right]^{-1} \frac{\partial \omega_{01}}{\partial l_{b}} \frac{\partial \omega_{01}}{\partial l_{b'}}  \q .
\end{equation}
Note that we have already proven that \eqref{eq:form_inv_1} holds due to the identity \eqref{eq:der_deficit_bdry}. This shows form invariance of the action.

For the invariance of the measure $\mu(l)$ in   (\ref{eq:3-2_pi}) we examine the contribution from the Gaussian integral:
\begin{equation} \label{eq:gaussian_contr_3-2}
\frac{\sqrt{2 \pi}}{\sqrt{\det (W_0)}} = \frac{\sqrt{2 \pi}}{\sqrt{H^{(3)}_{(01),(01)}}} = \frac{\sqrt{2 \pi}}{\sqrt{- \frac{\partial \omega_{01}}{\partial l_{01}}}} = \frac{\sqrt{12 \pi}}{l_{01}} \sqrt{\frac{ V_{\bar{2}} V_{\bar{3}} V_{\bar{4}}}{V_{\bar{0}} V_{\bar{1}}}}
\end{equation}
Hence, choosing the  measure factor as
\begin{equation}
\mu(l) = \frac{\prod_e \frac{l_e}{\sqrt{12 \pi}}}{\prod_\tau \sqrt{V_{\tau}}}
\end{equation}
we obtain a partition function, invariant under $3-2$ Pachner moves. 
Here $e$ denotes the edges and $\tau$ the tetrahedra in the triangulation.

\subsection{$4 -1$ move} \label{sec:com_4-1}

For 3D gravity, in addition to the $3-2$ move, we have also to consider the $4-1$ move. This move amounts to the subdivision of one tetrahedron, denoted by $(1 2 3 4)$, into four by adding one additional vertex $(0)$, placing it inside the original tetrahedron and connecting it with all of the remaining vertices, see section \ref{sec:Pachner_4-1}.

In contrast to the $3- 2$ move, the edge lengths of the new edges, i.e. the position of the new vertex inside the original tetrahedron, is not uniquely fixed. In fact the action is invariant under translations of the vertex $(0)$ inside the tetrahedron $(1 2 3 4)$, such that one expects the Hessian matrix to have three null eigenvectors. In order to compute this matrix, terms of the form $\frac{\partial \omega_e}{\partial l_{e'}}$ have to be evaluated just as in the $3-2$ move. Following a similar derivation as in the previous section, one arrives at the following terms:
\begin{itemize}
 \item In case either the edge $e=(ij)$ or edge $e'=(km)$ are in the bulk, one obtains:
  \begin{equation} \label{eq:4-1_bulk}
    \frac{\partial^2 S}{\partial l_{km} \partial l_{ij}} = (-1)^{s_{i} + s_j + s_{k} + s_m + 1} \frac{l_{ij} l_{km}}{6} \frac{V_{\bar{i}} V_{\bar{j}} V_{\bar{k}} V_{\bar{m}}}{\prod_{n} V_{\bar{n}}}
    \end{equation}
    where 
    \begin{equation}
     s_i = \left\{
     \begin{array}{l r}
     1 & \quad \text{if } i=0 \\
     0 & \quad \text{else}
     \end{array}
     \right. \q .
    \end{equation}
 \item In case both edges are in the boundary, one obtains:
  \begin{equation} \label{eq:4-1_bdry}
    \frac{\partial^2 S}{\partial l_{km} \partial l_{ij}} = (-1)^{s_{i} +s_j + s_{k} +s_m + 1} \frac{l_{ij} l_{km}}{6} \frac{ V_{\bar{i}} V_{\bar{j}} V_{\bar{k}} V_{\bar{m}}}{\prod_{n} V_{\bar{n}}} - \frac{\partial \omega^{(1)}_{ij}}{\partial l_{km}}
    \end{equation}
    where $\omega^{(1)}$ denotes an exterior dihedral angle in the one tetrahedron configuration.
\end{itemize}

Again, notice the simple form of the Hessian,
\ba\label{bia2101}
H^{(4)}_{(ij),(km)} = H^{(1)}_{(ij),(km)} + c \,h_{(ij)} \,h_{(km)} \q ,
\ea
with a factorizing summand. This form of the Hessian makes the appearance of null vectors obvious.


\subsubsection{Null eigenvectors} \label{sec:null_ev_3d}


Since the pure bulk part $H^{(4)}_{(0i),(0j)}$  of the Hessian matrix factorizes, we can easily examine the condition for null vectors $\vec{v}$:
\begin{equation}
\sum_{j} H_{(0i),(0j)} v_j = c h_{0i} \sum_{j} h_{0j} v_j \overset{!}{=} 0
\end{equation}
Hence, due to the factorizing form of the Hessian, we just have one condition for the null vectors. Therefore the Hessian has three null eigenvectors and of the four bulk degrees of freedom three are gauge.

Furthermore we have to discuss the sign of the Hessian. The only non--vanishing eigenvalue of the submatrix $H_{(0i),(0j)}$ can also easily be determined due to the factorizing form to be $\sum_j H_{(0j),(0j)}$. This gives a negative eigenvalue, hence the Gaussian integral would not be convergent. This is a trace of the conformal mode problem in Euclidean gravity: the kinematic term of the conformal mode comes with the `wrong' sign, so that the Euclidean action is not bounded from below. We see that the Pachner moves allow a nice isolation of this mode problem into the $4-1$ moves. We will change the global sign for the action of the $4-1$ move, such that the integral (over the one non-gauge mode) converges. This can be understood as selecting a complex contour for the integration for the conformal and the other modes separately, see \cite{Gibbons:1978ac} for a discussion in the continuum.


\subsubsection{Invariance of the path integral}

Similar to the $3 -2$ move, we have to consider
\begin{equation} \label{eq:4-1_pi}
P_{4\rightarrow 1} = \int \prod_{i} d \lambda_{0i}\, \mu(l) \; \exp \left\{ - \sum_{(ij),(km)} \frac{1}{2} H^{(4)}_{(ij),(km)} \, \lambda_{ij} \, \lambda_{km} \right \}
\end{equation}
where $\mu(l)$ is again a measure factor, which we assume to depend only on background variables $l$, which have to make up a flat configuration. \eqref{eq:4-1_pi} is again a partial Gaussian integral but with three gauge degrees of freedom, for which we will modify the general method of section \ref{sec:inv_pi_3-2}. 

Again the general form for the Gaussian integral is:
\begin{equation} 
I = \int dq_1 \ldots dq_r \exp \left \{- \frac{1}{2} \vec{q}^T M \vec{q} \right \}  \q .
\end{equation}
Since there are gauge degrees of freedom one integrates over the matrix $M$ is singular. Assume that there are $m$ gauge degrees of freedom 
such that we can split $\vec{q}$ in the following way:
\begin{equation}
\vec{q} = (\underbrace{q_1, \ldots, q_{r-m}}_{=:\vec{w}}, \underbrace{ q_{r-m+1}, \ldots, q_r}_{=:\vec{g}},\underbrace{ q_{r+1}, \ldots, q_n}_{=:\vec{u}}) \q .
\end{equation}
(Here we assume that  the transformation between $q_{r-m+1},\ldots, q_r$ and the $m$ gauge parameters is not singular.)
This implies the following split for the matrix $M$
\begin{equation}
M = \left(
\begin{array}{c c c}
\; W_0 \; & \; V_g \; & \; V \; \\
\; V_g^T \; & \; G_0 \; & \; Z_0 \; \\
\; V^T \; & \; Z_0^T \; & \; U_0 \;
\end{array}
\right)
\end{equation}
where $W_0$ is non-singular. Integrating out the degrees of freedom summarized in $\vec{w}$ one obtains:

\begin{equation}
I = \frac{(2 \pi)^{\frac{(r-m)}{2}}}{\sqrt{\det (W_0)}} \exp \left\{- \frac{1}{2} \left(\vec{g}^T G \vec{g} + \vec{g}^T Z \vec{u} + \vec{u}^T Z^T \vec{g} + \vec{u}^T U \vec{u} \right) \right\}
\end{equation}
with 
\begin{align}
G =& G_0 - V_g^T W_0^{-1} V_g  \q ,\\
Z = &Z_0 - V_g^T W_0^{-1} V \q ,\\
U =& U_0 - V^T W_0^{-1} V  \q .
\end{align}
Applying this formalism to the problem under discussion, one identifies (here $i,j\neq 0,1$):
\begin{xalignat}{2}
{(W_0)}_{(01),(01)} &= H^{(4)}_{(01),(01)}  = - \frac{\partial \omega_{01}}{\partial l_{01}}  \q, &
(G_0)_{(0i),(0j)}&= \frac{\partial \omega_{01}}{\partial l_{0i}} \frac{\partial l_{01}}{\partial l_{0j}}  \q, \nn\\
(U_0)_{b,b'} &= H^{(1)}_{b,b'} + \frac{\partial \omega_{01}}{\partial l_{b}} \frac{\partial l_{01}}{\partial l_{b'}}  \q ,&
(V_g)_{(01),(0i)} &= -\frac{\partial \omega_{01}}{\partial l_{0i}} \q , \nn\\
(V)_{(01),b} & = -\frac{\partial \omega_{01}}{\partial l_{b}} \q, &
(Z_0)_{(0i),b} &= \frac{\partial \omega_{01}}{\partial l_{0i}} \frac{\partial l_{0i}}{\partial l_{b}} \q .
\end{xalignat}

We therefore obtain:
\begin{align}
(G)_{(0i),(0j)} &=  \frac{\partial \omega_{01}}{\partial l_{0i}} \frac{\partial l_{01}}{\partial l_{0j}} + \left[\frac{\partial \omega_{01}}{\partial l_{01}} \right]^{-1} \frac{\partial \omega_{01}}{\partial l_{0i}} \frac{\partial \omega_{01}}{\partial l_{0j}} \overset{\eqref{eq:der_deficit_bdry}}{=} 0
\\
(Z)_{(0i),b}  &=  \frac{\partial \omega_{01}}{\partial l_{0i}} \frac{\partial l_{01}}{\partial l_{b}} + \left[\frac{\partial \omega_{01}}{\partial l_{01}} \right]^{-1} \frac{\partial \omega_{01}}{\partial l_{0i}} \frac{\partial \omega_{01}}{\partial l_{b}} \overset{\eqref{eq:der_deficit_bdry}}{=} 0
\\
(U)_{b,b'} &=H^{(1)}_{b,b'}  + \frac{\partial \omega_{01}}{\partial l_{b}} \frac{\partial l_{01}}{\partial l_{b'}} + \left[\frac{\partial \omega_{01}}{\partial l_{01}} \right]^{-1} \frac{\partial \omega_{01}}{\partial l_{b}} \frac{\partial \omega_{01}}{\partial l_{b'}} \overset{\eqref{eq:der_deficit_bdry}}{=} H^{(1)}_{b,b'} \q .
\end{align}
This proves form invariance of the action, as the remaining term in the exponential corresponds to the action of the tetrahedron $(1234)$ (after we have rotated back the global sign of the action).

Note that after having only integrated over $\lambda_{01}$ the other bulk variables $\lambda_{0i},\,i=2,3,4$ do not appear anymore in the exponential. 

 Let us first consider how the  measure factor is modified by the Gaussian integration over $\lambda_{01}$. The additional factor is given by
\begin{equation}
\frac{\sqrt{2 \pi}}{\sqrt{\det (W_0)}} = \frac{\sqrt{12 \pi}}{l_{01}} \sqrt{\frac{V_{\bar{2}} V_{\bar{3}} V_{\bar{4}}}{V_{\bar{0}} V_{\bar{1}}}} \q .
\end{equation}
If we consider $\mu(l)$  in (\ref{eq:4-1_pi}) to be the same measure which gives an invariant amplitude under the $3-2$ move, 
namely
\be
\mu(l)= \frac{\prod_e \frac{l_e}{\sqrt{12 \pi}}}{\prod_\tau \sqrt{V_{\tau}}}
\ee
where $e$ includes all boundary and bulk edges and $\tau$ the four tetrahedra of the initial configuration,
we obtain:
\begin{equation} \label{41result}
P_{4 - 1} = \frac{\prod_{b} l_{b}}{\sqrt{12 \pi V_{\bar{0}}}} \exp \left\{- \sum_{b,b'} \frac{1}{2} H^{(1)}_{b,b'}\, \lambda_{b} \, \lambda_{b'} \right\} \int \frac{\prod_{i\neq 1} \frac{l_{0i}}{\sqrt{12 \pi}} d \lambda_{0i}}{V_{\bar{1}}}  \q .
\end{equation}

The remaining integral over the variables $\lambda_{02},\ldots, \lambda_{04}$ can be identified with an integration over the gauge orbit, which is given by the displacement of the inner vertex $(0)$. As one can show \cite{Korepanov:2000aj,Baratin:2006yu}, see also appendix \ref{app:eucl_int}, the following identity between integration measures holds
\ba \label{eq:eucl_m_3d}
d^3x^\alpha_0=  \frac{\prod_{i=2,3,4} l_{0i} \,dl_{0i}}{ 6 V_{\bar{1}}}  \q .
\ea
Here $d^3x^\alpha_0$ is the integration measure of the Euclidean coordinates $x_0^\alpha,\, \alpha=1,2,3$ of the vertex $(0)$. The displacement of this vertex corresponds exactly to the gauge action of the discrete remnant of the diffeomorphisms \cite{hamber3d,Freidel:2002dw,dittrichreview}. Hence we will replace the last factor in (\ref{41result}) by $1$. This can be understood as resulting from a gauge fixing procedure, including the appropriate Faddeev-Popov determinant.  (The numerical factors are chosen to conform with the integration measure found for the $3-2$ move, however it is not possible to fix them uniquely.)

\section{Summary for 3D gravity} \label{sec:summary_3d}

For a general 3D triangulation we define the path integral for linearized Regge calculus  by
\begin{equation} \label{eq:path_int_3d}
P :=  \int     \frac{\prod_{e} \frac{l_e}{\sqrt{12 \pi}}}{\prod_\tau \sqrt{V_{\tau}}} \prod_{e \subset \text{bulk}} d \lambda_{e} \, \exp \left\{-\frac{1}{2} \sum_{e,e'} H_{e,e'} \lambda_e \lambda_{e'} \right\} \q .
\end{equation}
$l_e$ is the length of the edge $e$, $V_{\tau}$ is the volume of the tetrahedron $\tau$, $\lambda_e$ is the edge length perturbation of the edge $e$ and $H_{e,e'}$ is the $e$-$e'$ matrix element of the Hessian matrix of the Regge action. The considerations conducted in the previous section show that \eqref{eq:path_int_3d} is invariant under Pachner moves, in case one follows the gauge fixing and sign rotation procedure procedure for the $4-1$ move discussed above. Hence (\ref{eq:path_int_3d}) does not depend on the choice of bulk triangulation and in this sense it is discretization independent.

Here we assigned the numerical pre--factor $(12\pi)^{-1/2}$ to the edges of the triangulation (as the $\pi$ factors result from integration over edges). Another possibility would be to associate this pre--factor to the tetrahedra of the triangulation, in which case one needs to appropriately adjust the numerical constant in the gauge fixing prescription for the $4-1$ move.

Amazingly, the path integral measure which we found for linearized Regge calculus, coincides with the semi-classical limit of the Ponzano--Regge model \cite{pr,roberts,hellmannpr}. This is a triangulation independent\footnote{also requiring a gauge fixing procedure for the $4-1$ moves} spin foam model for 3D quantum gravity. Here the numerical pre--factors (also given by $(12\pi)^{-1/2}$) are associated to the tetrahedra. 

It would be interesting to see, whether this correspondence can be extended to 3D Regge calculus with a cosmological constant.  This theory can be (classically) formulated in a triangulation independent way, by using curved tetrahedra \cite{bahrdittrich2,newregge}. The corresponding quantization is given by the Turaev-Viro model \cite{tuarev-viro}, for which the semi--classical limit has been obtained \cite{robertslambda}. Hence (\ref{eq:path_int_3d}) should give a triangulation independent amplitude for linearized Regge calculus with a (positive) cosmological constant by replacing $l_e$ and $V_{\tau}$ by their respective counterparts on the sphere, i.e. $\sin(l_e)$, where $l_e \in [0,\pi]$, and $\mathcal{V}_{\tau}$, the determinant of the Gram matrix.

\section{Computation of the Hessian matrix in 4D} \label{sec:Hessian_4d}

We are now going to discuss the 4D case.  We will proceed as for 3D, that is first determine the matrix elements of the Hessian and then consider the path integral for the Pachner moves. It will turn out that the $4-2$ and $5-1$ moves behave very similarly to the $3-2$ and the $4-1$ move, respectively, in 3D. There is however an additional Pachner move in 4D, namely the $3-3$, which is significantly different, and thus responsible for the non--trivial dynamics of 4D Regge gravity.

The Hessian of the Regge action is given by
\begin{equation} \label{eq:formula_second_der}
 \frac{\partial^2 S}{\partial l_{ij} \partial l_{mn}} = -\sum_{stu} \frac{\partial A_{stu}}{\partial l_{mn}} \frac{\partial \omega_{stu}}{\partial l_{ij}} - \sum_{stu \, \in \text{bulk}} \frac{\partial^2 A_{stu}}{\partial l_{ij} \partial l_{mn}} \omega^{(bulk)}_{stu} - \sum_{stu \, \in \text{bdry}} \frac{\partial^2 A_{stu}}{\partial l_{ij} \partial l_{mn}} \omega^{(bdry)}_{stu}
\end{equation}
where $\omega_{stu}$ is the deficit angle at the (bulk or boundary) triangle $(stu)$. In the following we will not discuss the last two terms in \eqref{eq:formula_second_der} because we will consider flat background solutions, i.e. $\omega^{(bulk)}=0$. 
That is the second term in (\ref{eq:formula_second_der}) vanishes and the third term is unaffected by Pachner moves, since the extrinsic geometry, defined by the embedding into flat space, is not changed. (Furthermore this term is only multiplied by boundary perturbations, which are not integrated over in the path integrals.)

Hence we define a reduced Hessian matrix:
\begin{equation}
H_{(ij),(km)}:=- \sum_{\Delta} \frac{\partial A_{\Delta}}{\partial l_{ij}} \frac{\partial \omega_{\Delta}}{\partial l_{km}}
\end{equation}
which can be rewritten as the product of the following two matrices:
\begin{equation}
H =- \left(
\begin{array}{c c c}
\; \frac{\partial A_{\Delta_1}}{\partial l_{01}} \; & \; \frac{\partial A_{\Delta_2}}{\partial l_{01}} \; & \ldots \\
\; \frac{\partial A_{\Delta_1}}{\partial l_{02}} \;& \ddots &  \\
\vdots & & 
\end{array}
\right)
\left(
\begin{array}{c c c}
\; \frac{\partial \omega_{\Delta_1}}{\partial l_{01}} \; & \; \frac{\partial \omega_{\Delta_1}}{\partial l_{02}} \; & \ldots \\
\; \frac{\partial \omega_{\Delta_2}}{\partial l_{01}} \;& \ddots & \\
\vdots & & 
\end{array} \q .
\right)
\end{equation}
Here the index summation is over all triangles in the triangulation.
So as in 3D we have to compute terms of the form $\frac{\partial \omega_{\Delta}}{\partial l_e}$. To this end we will proceed similarly as in the 3D case, as described in the next sections.

\subsection{$4- 2$ move} \label{sec:der_4-2}

\subsubsection{Auxiliary formulas} \label{sec:aux_form_4d}

As in 3D we will need additional formulas to derive all entries of the Hessian matrix in a compact way. We start with the derivatives of the dihedral angles at a given triangle with respect to the length of the opposite edge \eqref{eq:der_opposite}
\begin{equation*}
\frac{\partial \theta^{(ijkmn)}_{ijk}}{\partial l_{mn}} = \frac{l_{mn} A_{ijk}}{12 V}
\end{equation*}
where $A_{ijk}$ denotes the area of the triangle $(ijk)$, $l_{mn}$ is the lengths of the edge $(mn)$ and $V$ is the volume of the 4-simplex $(i j k m n)$. 

Now given a flat triangulation with six, i.e. $D+2$, vertices, we will consider edge lengths variations of at least two edges under the condition that the triagulation remains flat, i.e. the deficit angles are vanishing, $\omega=0$. Then, along the same line of arguments as in 3D, we obtain \cite{Korepanov:2002tp}:
\begin{itemize}
 \item In case the varied edges share a vertex:
 \begin{equation} \label{eq:edge_rel1_4d}
 \left| \frac{\partial l_{ij}}{\partial l_{jk}} \right| = \left| \frac{l_{jk}}{l_{ij}}\, \frac{V_{\bar{k}}}{V_{\bar{i}}} \right|
 \end{equation}
 \item In case the varied edges do not share a vertex:
 \begin{equation} \label{eq:edge_rel2_4d}
 \left|\frac{\partial l_{ij}}{\partial l_{km}} \right| = \left| \frac{l_{km}}{l_{ij}}\,\frac{V_{\bar{k}} V_{\bar{m}}}{V_{\bar{i}} V_{\bar{j}}} \right|
 \end{equation}
\end{itemize}
where $V_{\bar{k}}$ denotes the volume of the 4-simplex formed without the vertex $k$. To determine the sign of the derivatives in  (\ref{eq:edge_rel1_4d},\ref{eq:edge_rel2_4d}) one has to consider the geometric set--up  in detail.
Note that \eqref{eq:edge_rel1_4d} and \eqref{eq:edge_rel2_4d} are the exact 4D analogues of \eqref{eq:edge_rel1} and \eqref{eq:edge_rel2} respectively.

\subsubsection{Computation of $\frac{\partial \omega}{\partial l}$}

Consider the $4-2$ move, that is  two 4-simplices $(02345)$ and $(12345)$, which share one common tetrahedron $(2345)$. By connecting vertices $(0)$ and $(1)$, the two 4-simplices are split into four, namely $(01ijk)$.  This edge is the only bulk edge in the configuration with $4$ simplices. As for the $3-2$ move we can use two facts,  to specify the matrix elements of the Hessian. Namely on the one hand, that the equations of motions for the perturbation variable $\lambda_{01}$ require flatness, on the other hand that lengths perturbations around flat space have to satisfy equations (\ref{eq:edge_rel1_4d},\ref{eq:edge_rel2_4d}). The equation of motion is given by
\ba
\sum_{e\neq (01)}H_{(01),e}\lambda_e + H_{(01)(01)}\lambda_{01}=0  \q .
\ea
But as the perturbative solutions are also flat, the perturbation variables $\lambda_e$ have to satisfy the relations  (\ref{eq:edge_rel1_4d},\ref{eq:edge_rel2_4d}). Hence considering boundary data, such that only one $\lambda_e \neq 0$ for $e=(km)\neq(01)$  and $\lambda_{01}\neq 0$ we can deduce
 \begin{equation} \label{bia2100}
\left| \frac{ H_{(01),(km)}}{H_{(01),(01)}}\right| =
 \left|\frac{\delta l_{01}}{\delta l_{km}} \right| = \left| \frac{l_{km}}{l_{01}}\,\frac{V_{\bar{k}} V_{\bar{m}}}{V_{\bar{0}} V_{\bar{1}}} \right|  \q .
 \end{equation}
 
 To specify the (reduced) Hessian even further, we can use that the linearized deficit angles also have to vanish. That is consider variations of two edge lengths, here $\delta l_{01}=\lambda_{01}$ and $\delta l_{34}=\lambda_{34}$. Using that the linearized deficit angle $\delta\omega_{012}$ has to vanish we obtain
%
%
\begin{align}
0 = \delta \omega_{012} = & -\frac{\partial \theta_{012}^{(01234)}}{\partial l_{34}} \delta l_{34} + \frac{\partial \omega_{012}}{\partial l_{01}} \delta l_{01} \\
\overset{\eqref{eq:der_opposite}}{=} & - \frac{l_{34} A_{012}}{12 V_{\bar{5}}} \delta l_{34} + \frac{\partial \omega_{012}}{\partial l_{01}} \delta l_{01} 
\end{align}
\begin{equation}
\implies \left| \frac{\partial \omega_{012}}{\partial l_{01}} \right| = \left| \frac{l_{01} A_{012}}{12 V_{\bar{5}}} \frac{\delta l_{34}}{\delta l_{01}} \right| \overset{\eqref{eq:edge_rel2_4d}}{=} \left|\frac{l_{01} A_{012}}{12} \frac{V_{\bar{0}} V_{\bar{1}}}{V_{\bar{3}} V_{\bar{4}} V_{\bar{5}}} \right|  \q .
\end{equation}
This can be repeated for all bulk deficit angles:
\begin{equation}
\left| \frac{\partial \omega_{01i}}{\partial l_{01}} \right| = \left|\frac{l_{01} A_{01i}}{12} \frac{V_{\bar{0}}^2 V_{\bar{1}}^2 V_{\bar{i}}}{\prod_{n} V_{\bar{n}}} \right| \q .
\end{equation}
The sign is again determined by the geometry under consideration.

The rest of the derivations proceeds in the same way as for the $3-2$ move. That is for the derivation of the bulk deficit angle with respect to a boundary edge length $l_b$ we obtain
\begin{equation} \label{2201}
\frac{\partial \omega_{01i}}{\partial l_{b}} = -\frac{\partial \omega_{01i}}{\partial l_{01}} \frac{\partial l_{01}}{\partial l_{b}} \q .
\end{equation}

To determine the derivatives of the (boundary) exterior angles one again uses that these angles agree in both configurations of the $4-2$ move. This results in
%
%
\begin{equation}
\left| \frac{\partial \omega^{(4)}_{ijk}}{\partial l_{01}} \right| = \left| \frac{l_{01} A_{ijk}}{12} \frac{V_{\bar{0}} V_{\bar{1}} V_{\bar{i}} V_{\bar{j}} V_{\bar{k}}}{\prod_{n} V_{\bar{n}}} \right|
\end{equation}
and furthermore in
\begin{equation} \label{eq:der_deficit_bdry_4d}
 \frac{\partial \omega^{(4)}_{ijk}}{\partial l_{b}} = \frac{\partial \omega^{(2)}_{ijk}}{\partial l_{b}} - \frac{\partial \omega^{(4)}_{ijk}}{\partial l_{01}} \frac{\partial l_{01}}{\partial l_{b}}  \q .
\end{equation}
The missing signs are dependent on the geometry under discussion and determined by similar considerations as in section \ref{sec:sign_3d}. To summarize the results for the $4 - 2$ move:
\begin{itemize}
\item In the case that either the triangle or the edge is in the bulk:
 \begin{equation} 
  \frac{\partial \omega^{(4)}_{ijk}}{\partial l_{mn}} = (-1)^{s_{i} + s_j +s_k + s_{m} +s_n + 1} \frac{l_{mn} A_{ijk}}{12} \frac{V_{\bar{i}} V_{\bar{j}} V_{\bar{k}} V_{\bar{m}} V_{\bar{n}}}{\prod_{p} V_{\bar{p}}}
 \end{equation}
 where
 \begin{equation}
  s_i = \left \{
  \begin{array}{l r}
   1 & \quad \text{if } i \in \{0,1\} \\
   0 & \quad \text{else}
  \end{array}
  \right. .
 \end{equation}
\item In the case that both triangle and edge are in the boundary:
 \begin{equation}
  \frac{\partial \omega^{(4)}_{ijk}}{\partial l_{mn}} = (-1)^{s_{i} + s_j +s_k + s_{m} +s_n +1} \frac{l_{mn} A_{ijk}}{12} \frac{V_{\bar{i}} V_{\bar{j}} V_{\bar{k}} V_{\bar{m}} V_{\bar{n}}}{\prod_{p} V_{\bar{p}}} + \frac{\partial \omega^{(2)}_{ijk}}{\partial l_{mn}}
 \end{equation}
\end{itemize}
Note that as in 3D the formulas for $\frac {\partial \omega}{\partial l}$ factorize.

\subsubsection{Hessian matrix}

In order to complete the calculation for the (reduced) Hessian matrix, the terms $\frac{\partial A_{\Delta}}{\partial l_{ij}} \frac{\partial \omega_{\Delta}}{\partial l_{km}}$ have to be summed up, where 
\begin{equation} \label{eq:der_area}
\frac{\partial A_{ijk}}{\partial l_{ij}} = \frac{l_{ij}}{8 A_{ijk}} \underbrace{(l_{ik}^2 + l_{jk}^2 - l_{ij}^2)}_{=:F_{ij;k}} .
\end{equation}
Note that \eqref{eq:der_area} is only non-vanishing for four triangles in the triangulation for a given edge $(ij)$. This implies (in case either $(op)$ or $(mn)$ are in the bulk):
\begin{align} \label{bianeed}
H^{(4)}_{(op),(mn)}
 =& - \sum_{(ijk)} \frac{\partial A_{ijk}}{\partial l_{op}} \frac{\partial \omega_{ijk}}{\partial l_{mn}} = - \sum_{k\neq o,p} \frac{\partial A_{opk}}{\partial l_{op}} \frac{\partial \omega_{opk}}{\partial l_{mn}} \\
=& \sum_{k\neq o,p} \frac{1}{8} \frac{l_{op} l_{mn}}{12} (-1)^{s_{o} + s_p + s_k + s_{m} + s_n} \frac{V_{\bar{o}} V_{\bar{p}} V_{\bar{m}} V_{\bar{n}}}{\prod_{l} V_{\bar{l}}} V_{\bar{k}} F_{op;k} \nonumber \\
=& \underbrace{(-1)^{s_o + s_p + s_{m}+ s_n} \frac{l_{op} l_{mn}}{96} \frac{V_{\bar{o}} V_{\bar{p}} V_{\bar{m}} V_{\bar{n}}}{\prod_{l} V_{\bar{l}}}}_{\text{symmetric in $(op) \leftrightarrow (mn)$}} \underbrace{\sum_{k \neq o,p} (-1)^{s_{k}} V_{\bar{k}} F_{op;k}}_{=:D_{op}}
\end{align}
where $D_{op}$ is a factor independent of the choice of $(mn)$. In case $(op)$ and $(mn)$ are in the boundary, we obtain:
\begin{equation}
H^{(4)}_{(op),(mn)}
= D_{op} \, (-1)^{s_o + s_p + s_{m}+ s_n} \frac{l_{op} l_{mn}}{96} \frac{V_{\bar{o}} V_{\bar{p}} V_{\bar{m}} V_{\bar{n}}}{\prod_{l} V_{\bar{l}}} \underbrace{- \sum_{k \neq o,p} \frac{l_{op}}{8 A_{opk}} F_{op;k} \frac{\partial \omega^{(2)}_{opk}}{\partial l_{mn}}}_{=H^{(2)}_{(op),(mn)}} \q .
\end{equation}
$H^{(4)}$ and $H^{(2)}$ are equal to matrices of second derivatives of the Regge action up to symmetric terms. Hence $H^{(4)}$ and $H^{(2)}$ are also 
symmetric matrices. From this it follows that the factor $D_{op}$ is the same for all choices of $op$:
\begin{equation}
D:=D_{op} = D_{mn}  \q .
\end{equation}
Note that  the elements $H^{4}_{(ij),(01)}$ satisfy the relation (\ref{bia2100}).

\subsubsection{Invariance of the path integral}

We have to consider:
\begin{equation} \label{eq:4-2_pi}
P_{4 - 2} = \int d \lambda_{01} \, \mu(l) \, \exp \left\{- \sum_{(ij),(km)} \frac{1}{2} H^{(4)}_{(ij),(km)} \, \lambda_{ij} \, \lambda_{km} \right\}
\end{equation}
where the measure factor $\mu(l)$ is supposed to depend only on the background edge lengths $l$. Note that $H^{(4)}_{(01),(01)}>0$ such that \eqref{eq:4-2_pi} converges. (Note that in (\ref{eq:4-2_pi}) we did not include boundary terms which only depend on the boundary perturbations.)

The computation is analogous to section \ref{sec:inv_pi_3-2}. That is to show (form) invariance of the action, i. e.\ 
\begin{equation}
P_{4 - 2} \propto \exp \left\{ - \sum_{(ij)\neq (01),(km) \neq (01)}\frac{1}{2} \, H^{(2)}_{(ij)(km)}\, \lambda_{ij} \, \lambda_{km} \right \}
\end{equation}
we have to proof that
\begin{align} 
&H^{(4)}_{(ij),(mn)} = \left[H^{(4)}_{(01),(01)}\right]^{-1} H^{(4)}_{(ij),(01)} H^{(4)}_{(01),(mn)} \, + \, H^{(2)}_{(ij),(mn)} \nonumber \\
\overset{\eqref{eq:der_deficit_bdry_4d}}{\iff}&- \sum_{k \neq i,j} \frac{\partial A_{ijk}}{\partial l_{ij}} \frac{\partial \omega_{ijk}}{\partial l_{01}} \frac{\partial l_{01}}{\partial l_{mn}} \nonumber \\
=&\left[ \sum_{k \neq 0,1} \frac{\partial A_{01k}}{\partial l_{01}} \frac{\partial \omega_{01}}{\partial l_{01}} \right]^{-1} \left(\sum_{k \neq i,j} \frac{\partial A_{ijk}}{\partial l_{ij}} \frac{\partial \omega_{ijk}}{\partial l_{01}} \right) \left( \sum_{k\neq 0,1} \frac{\partial A_{01k}}{\partial l_{01}} \frac{\partial \omega_{01k}}{\partial l_{mn}} \right) \label{eq:form_inv_1_4-2} \q .
\end{align}
$(ij)$ and $(mn)$ denote two boundary edges. Applying \eqref{eq:der_deficit_bdry_4d} to the last term on the right hand side of \eqref{eq:form_inv_1_4-2} gives:
\begin{equation}
\left(\sum_{k \neq 0,1} \frac{\partial A_{01k}}{\partial l_{01}} \frac{\partial \omega_{01k}}{\partial l_{mn}} \right) \overset{\eqref{eq:der_deficit_bdry_4d}}{=} - \left( \sum_{k \neq 0,1} \frac{\partial A_{01k}}{\partial l_{01}} \frac{\partial \omega_{01k}}{\partial l_{01}} \right) \frac{\partial l_{01}}{\partial l_{mn}} \q , 
\end{equation}
which shows that \eqref{eq:form_inv_1_4-2} holds.

For the measure factor we examine the contribution from the Gaussian integral:
\begin{equation} \label{eq:gaussian_contr_4-2}
\frac{\sqrt{2 \pi}}{\sqrt{H^{(4)}_{(01),(01)}}} = \frac{\sqrt{2 \pi}}{\sqrt{- \sum_{k\neq 0,1} \frac{\partial A_{01k}}{\partial l_{01}} \frac{\partial \omega_{01k}}{\partial l_{01}}}} = \frac{\sqrt{192 \pi}}{l_{01}} \sqrt{\frac{ V_{\bar{2}} V_{\bar{3}} V_{\bar{4}} V_{\bar{5}}}{V_{\bar{0}} V_{\bar{1}}}} \frac{1}{\sqrt{D}}
\end{equation}
Apart from the additional factor $\frac{1}{\sqrt{D}}$, \eqref{eq:gaussian_contr_4-2} is of a similar form as \eqref{eq:gaussian_contr_3-2} for the $3 - 2$ move in 3D. Hence the invariant measure factor $\mu(l)$ should be proportional to:
\begin{equation}
\mu(l) = \frac{\prod_e \frac{l_e}{\sqrt{192 \pi}}}{\prod_\Delta \sqrt{ V_{\Delta}}}
\end{equation}
where $e$ denotes the edges and $\Delta$ the 4--simplices in the triangulation.  However, with this form, we will still get factors of  $\frac{1}{\sqrt{D}}$ by applying $4-2$ Pachner moves. The factor $D$ does not factorize into contributions that could be associated to 4--simplices or other subsimplices. It is rather a sum of terms involving the edge lengths of the entire triangulation associated to the $4-2$ move. We will therefore defer the discussion of this factor $D$ until after we have considered all the Pachner moves in 4D.

\subsection{$5- 1$ move}

Let us now consider the $5 -1$ move in 4D. Again, many derivations will be similar to the ones for the $4-1$ move in 3D.
The $5-1$ move corresponds to the subdivision of one 4--simplex, denoted by $(1 2 3 4 5)$, into five by adding one additional vertex $(0)$, placing it inside the original 4--simplex and connecting it with all of the remaining vertices, see section \ref{sec:Pachner_5-1}.

Here, the edge lengths of the new edges, i.e. the position of the new vertex inside the original 4--simplex, are not uniquely fixed. 
Accordingly there is a 4--parameter set of solutions and we expect to find four null modes in the Hessian.


The derivation of the matrix elements for the Hessian proceeds as for the $4 - 2$ move. We arrive at the following terms: 
\begin{itemize}
 \item In case either the edge $e=(op)$ or edge $e'=(mn)$ are in the bulk, one obtains:
  \begin{equation} \label{eq:5-1_bulk}
H_{(mn),(op)}^{(5)}= (-1)^{s_{o} + s_p + s_{m} + s_n + 1} \frac{l_{op} l_{mn}}{96} \frac{V_{\bar{o}} V_{\bar{p}} V_{\bar{m}} V_{\bar{n}}}{\prod_{l} V_{\bar{l}}} \underbrace{\sum_{k\neq o,p} (-1)^{s_k} F_{op;k} V_{\bar{k}}}_{:=D^{(5)}}
  \end{equation}
  where 
  \begin{equation}
   s_i = \left\{
   \begin{array}{l r}
   1 & \quad \text{if } i=0 \\
   0 & \quad \text{else}
   \end{array}
   \right.
  \end{equation}
  and $F_{op;k}$ is defined as in the previous section. Note, that as for the $4-2$ move the factor $D^{(5)}$ does not depend on the choice of indices $(op)$ in (\ref{eq:5-1_bulk}).
 \item In case both edges are in the boundary, one obtains:
  \begin{equation} \label{eq:5-1_bdry}
H_{(mn),(op)}^{(5)}= (-1)^{s_{o} +s_p + s_{m} +s_n + 1} \frac{l_{op} l_{mn}}{96} \frac{ V_{\bar{o}} V_{\bar{p}} V_{\bar{m}} V_{\bar{n}}}{\prod_{l} V_{\bar{l}}} D^{(5)} + H^{(1)}_{(op),(mn)}
  \end{equation}
  where $H^{(1)}_{(op),(mn)}$ denotes the $(op)$-$(mn)$ matrix element of the Hessian of the one 4--simplex configuration.
\end{itemize}
This gives all matrix elements of the Hessian of the $5-1$ move. In the next section we will discuss the pure bulk terms more closely, in particular with respect to null eigenvectors.

\subsubsection{Null eigenvectors}

In this section we examine the pure bulk terms of the Hessian matrix, i.e. equation \eqref{eq:5-1_bulk} for edges $(0i)$ and $(0j)$ for arbitrary $i,j$. Then \eqref{eq:5-1_bulk} can be rewritten as:
\begin{equation} \label{eq:5-1_bulk_bulk}
H^{(5)}_{(0i),(0j)}  = - \frac{l_{0i} l_{0j}}{96} \frac{V_{\bar{0}}^2 V_{\bar{i}} V_{\bar{j}}}{\prod_{l} V_{\bar{l}}} D^{(5)} =  \underbrace{l_{0i} V_{\bar{i}}}_{h_{0i}} \, \underbrace{l_{0j} V_{\bar{j}}}_{h_{0j}} \underbrace{(-1) \, D^{(5)} \, \frac{V_{\bar{0}}}{96 V_{\bar{1}} V_{\bar{2}} V_{\bar{3}} V_{\bar{4}} V_{\bar{5}}}}_{c}
\end{equation}
So, as in 3D, the bulk terms in the Hessian $H^{(5)}_{(0i),(0j)}$ factorize:
\begin{equation}
H^{(5)}_{(0i),(0j)} = c\, h_{0i} h_{0j}  \q .
\end{equation}

Hence, following the argument in section \ref{sec:null_ev_3d}, we can conclude that $H^{(5)}_{(0i),(0j)}$ features four null vectors. The one non-vanishing eigenvalue is again given by $\sum_j  H_{(0j),(0j)}$, which amounts to a negative value. We will proceed as for the $4-1$ move and change the global sign for the action associated to the $5-1$ move. This can again be interpreted as taking care of the conformal factor problem in Euclidean gravity \cite{Gibbons:1978ac}.



\subsubsection{Invariance of the path integral}

Similar to the $4- 2$ move, we have to consider
\begin{equation} \label{eq:5-1_pi}
P_{5- 1} = \int \prod_{i} d \lambda_{0i}\, \mu(l) \; \exp \left\{ - \sum_{(ij),(km)} \frac{1}{2} H^{(4)}_{(ij),(km)} \, \lambda_{ij} \, \lambda_{km} \right \}
\end{equation}
where $\mu(l)$ is a measure factor, which we assume to depend only on the background variables $l$, making up a flat configuration. \eqref{eq:5-1_pi} is again a partial Gaussian integral but with four gauge degrees of freedom. The treatment of this integral will be completely analogous to the $4-1$ move. Due to the gauge modes we will first integrate only over $\lambda_{01}$. 

This integration will result in an exponential that is independent of the other variables $\lambda_{0i},\,i=2,3,4,5$. To show that we obtain the action associated to the one remaining simplex $(12345)$ we need to invoke the identities
\begin{align}
& H^{(5)}_{(ij),(mn)} = \left[H^{(5)}_{(01),(01)} \right]^{-1} H^{(5)}_{(ij),(01)} H^{(5)}_{(01),(mn)} \, + \, H^{(1)}_{(ij),(mn)} \nonumber \\
\overset{\eqref{eq:der_deficit_bdry_4d}}{\iff} & - \sum_{k \neq i,j} \frac{\partial A_{ijk}}{\partial l_{ij}} \frac{\partial \omega_{ijk}}{\partial l_{01}} \frac{\partial l_{01}}{\partial l_{mn}} \nonumber \\
=&\left[\sum_{k \neq 0,1} \frac{\partial A_{01k}}{\partial l_{01}} \frac{\partial \omega_{01k}}{\partial l_{01}} \right]^{-1} \left( \sum_{k \neq i,j} \frac{\partial A_{ijk}}{\partial l_{ij}} \frac{\partial \omega_{ijk}}{\partial l_{01}} \right) \left( \sum_{k \neq 0,1} \frac{\partial A_{01k}}{\partial l_{01}} \frac{\partial \omega_{01k}}{\partial l_{mn}} \right) \label{eq:form_inv_1_5-1}
\end{align}
Similar to (\ref{eq:form_inv_1_4-2}), we apply (\ref{eq:der_deficit_bdry_4d}) to the last term in (\ref{eq:form_inv_1_5-1}), which gives:
\begin{equation}
\sum_{k \neq 0,1} \frac{\partial A_{01k}}{\partial l_{01}} \frac{\partial \omega_{01k}}{\partial l_{mn}} = - \sum_{k \neq 0,1} \frac{\partial A_{01k}}{\partial l_{01}} \frac{\partial \omega_{01k}}{\partial l_{01}} \frac{\partial l_{01}}{\partial l_{mn}} \q .
\end{equation}
This proves (\ref{eq:form_inv_1_5-1}) and hence form invariance of the linearized action under the $5-1$ move.

The measure factor $\mu(l)$ in (\ref{eq:5-1_pi}) changes by the $\lambda_{01}$ integration by a factor
\begin{equation}
\frac{\sqrt{2 \pi}}{\sqrt{ H_{(01),(01)}}} = \frac{\sqrt{192 \pi}}{l_{01}} \sqrt{\frac{V_{\bar{2}} V_{\bar{3}} V_{\bar{4}} V_{\bar{5}}}{V_{\bar{0}} V_{\bar{1}}}} \frac{1}{\sqrt{D^{(5)}}}
\end{equation}
which turns out to be of a similar form as the contribution in the $4 - 2$ move. If we consider $\mu(l)$ in (\ref{eq:5-1_pi}) to be given by
\be
\mu(l)= \frac{\prod_e \frac{l_e}{\sqrt{192 \pi}}}{\prod_\Delta \sqrt{V_{\Delta}}}
\ee
we obtain for the path integral 
\begin{equation} \label{51result}
P_{5-1} = \frac{\prod_{b} \frac{l_{b}}{\sqrt{192 \pi}}}{\sqrt{V_{\bar{0}}}} \frac{1}{\sqrt{D^{(5)}}} \exp \left\{- \sum_{b,b'} \frac{1}{2} H^{(1)}_{b,b'}\, \lambda_{b} \, \lambda_{b'} \right\} \int \frac{\prod_{i\neq 1} \frac{l_{0i}}{\sqrt{192 \pi}} d \lambda_{0i}}{ V_{\bar{1}}}
\end{equation}
 Also here the remaining integral can be identified with an integration over the gauge orbit, which is again given by the displacement of the inner vertex $(0)$. In 4D we have the identity
 \ba \label{eq:eucl_m_4d}
 d^4 x_0^\alpha=\frac{\prod_{i=2,3,4,5} l_{0i}\,dl_{0i}}{24 V_{\bar{1}}}
 \ea
where $d^4 x_0^\alpha$ is the integration measure for the Euclidean coordinates $x_0^\alpha,\,\alpha=1,2,3,4$ of the vertex $(0)$. Hence we will replace the remaining integral in  (\ref{51result}) by $1$.

\subsection{$3 - 3$ move}

In addition to the $5-1$ and $4-2$ moves, the set of Pachner moves in 4D includes the $3-3$ move. Here a complex of three 4--simplices is replaced with another complex of 3 4--simplices, such that the boundary triangulation is not changed. This move does not involve any bulk edge, hence, differing from all the other moves considered so far, we do not have an equation of motion associated to this move.

There is another essential difference to the other Pachner moves, namely that the action is not invariant under $3-3$ moves. Evaluating the (full) Regge action for the two configurations of the $3-3$ move, one finds a difference, that grows quadratically with the deficit angle of the (only) bulk triangle \cite{bahrprivate}. Hence the action is not invariant in general under 3-3 moves, in fact, such an invariance applies only on flat configurations.  This violation of the invariance of the action holds also for the quadratic action of the linearized theory, as can be expected from the behavior in the full theory and as can be checked explicitly on configurations with non--vanishing (linearized) curvature.

The derivation of the (reduced) Hessian matrix for the $3-3$ move proceeds in a slightly different way, as we now have to take into account that the boundary perturbations might describe curvature. The result will however have the same structure as for the other Pachner moves. 

To start the derivation note that in both configurations $A$ (with simplices $(01234)$, $(01235)$, $(01245)$) and $B$ (with simplices $(01345)$, $(02345)$, $(12345)$ there is only one bulk triangle, namely $(012)$ and $(345)$ respectively in $A$ and $B$. The vanishing of the linearized deficit angles defines boundary perturbations in flat directions:
\ba\label{331}
\sum_{(ij)}\frac{\partial \omega_{012}^A}{\partial l_{ij}}\lambda_{ij} =0= \sum_{(ij)}\frac{\partial \omega_{345}^B}{\partial l_{ij}}\lambda_{ij}  \q .
\ea

For such flat variations $\lambda_{ij}$ we can again derive, along the same arguments as in section \ref{sec:aux_form_4d}, the relations (\ref{eq:edge_rel2_4d})
 \begin{equation} \label{332}
 \left|\frac{\lambda_{ij}}{\lambda_{km}} \right| = \left| \frac{l_{km}}{l_{ij}}\,\frac{V_{\bar{k}} V_{\bar{m}}}{V_{\bar{i}} V_{\bar{j}}} \right| \q .
 \end{equation}
Note that equation (\ref{331}) also implies that the gradients of the deficit angles in the two configurations are parallel to each other. That is the space of flat boundary perturbations $\lambda_{ij}$ in both configurations is the same, the linearized curvature will however  have different values in the general case, if evaluated on the same set of (non--flat) boundary perturbations $\lambda_{ij}$. 

Now, starting with the derivatives
\ba\label{333}
\frac{\partial \omega_{012}^A}{\partial l_{34}}=-\frac{\partial \theta^{01234}_{012}}{\partial l_{34}}= -\frac{l_{45} A_{123}}{12 V_{\bar{5}}} \q ,\q\q
\frac{\partial \omega_{345}^B}{\partial l_{12}}=-\frac{\partial \theta^{12345}_{345}}{\partial l_{12}}= -\frac{l_{12} A_{345}}{12 V_{\bar{0}}} \q 
\ea
and using (\ref{331}) and (\ref{332}), assuming that only two appropriately chosen length perturbations $\lambda_{ij},\lambda_{mn}$ are not vanishing, we can obtain all other derivatives of the bulk deficit angle in both configurations. The signs can again be determined as in section \ref{sec:sign_3d}.

The result is given by
\ba\label{334}
\frac{\partial \omega_{012}^A}{\partial l_{ij}} &=& (-1)^{s^A_i + s^A_j + 1}\,  \frac{l_{ij} A_{012}} {12}\, \frac{V_{\bar{0}} V_{\bar{1}} V_{\bar{2}} V_{\bar{i}} V_{\bar{j}}}{\prod_p V_{\bar{p}}}  \nn\\
\frac{\partial \omega_{345}^B}{\partial l_{ij}} &=& (-1)^{s^B_i + s^B_j +1} \, \frac{l_{ij} A_{345}} {12}\, \frac{V_{\bar{3}} V_{\bar{4}} V_{\bar{5}} V_{\bar{i}} V_{\bar{j}}}{\prod_p V_{\bar{p}}} 
\ea
Here we defined the sign factors as
\ba\label{335}
s^A_i=\left\{ 
\begin{array}{l r}
1 & \q \text{if }i \in \{0,1,2\} \\
0 & \q \text{else}
\end{array} \right.
 \q,\q\q  s^B_i=\left \{
 \begin{array}{l r}
 1 & \q \text{if }i \in \{3,4,5\} \\
 0 & \q \text{else}
 \end{array}
 \right. \q .
\ea

With the understanding that $\omega_{345}^A\equiv 0=\omega_{012}^B$ we can write the relation between the derivatives as 
\ba\label{336}
\frac{\partial \omega_{opk}^A}{\partial l_{ij}} -\frac{\partial \omega_{opk}^B}{\partial l_{ij}} &=&
 \underbrace{(-1)^{s^A_i + s^A_j + s^A_o + s^A_p + s^A_k}}_{=(-1)^{s^B_i + \ldots + s^B_k +1}}\,  \frac{l_{ij} A_{opk}} {12}\, \frac{V_{\bar{o}} V_{\bar{p}} V_{\bar{k}} V_{\bar{i}} V_{\bar{j}}}{\prod_p V_{\bar{p}}}  
\ea
where $(opk)$ is either the set $(012)$ or $(345)$. 
Note that \eqref{336} is consistent with \eqref{335} under change $A \leftrightarrow B$, i.e. change of sign.

We will soon discover that (\ref{336}) holds also for the other (boundary) angles. 
To this end we use that the linearized boundary extrinsic curvature angles coincide in both configurations $A$ and $B$, if evaluated on flat boundary perturbations.  Hence we can conclude that the difference of the gradients of a given boundary angle has to be proportional to the gradient of one of the bulk angles, i.\ e.\ 
\ba
\frac{\partial \omega_{mnl}^A}{\partial l_{ij}} -\frac{\partial \omega_{mnl}^B}{\partial l_{ij}} = c^A_{mnl} \frac{\partial \omega_{012}^A}{\partial l_{ij}}  \q .
\ea
Again we can start with an especially simple derivative, i.e.
\ba
\frac{\partial \omega^A_{345}}{\partial l_{12}}= -\frac{l_{12} A_{345} }{12}\, \frac{V_{\bar{3}} V_{\bar{4}} V_{\bar{5}} V_{\bar{1}} V_{\bar{2}}}{\prod_p V_{\bar{p}}}  \q ,\q\q     \frac{\partial \omega^B_{345}}{\partial l_{12}}= 0 
\ea
to get hold of all the other derivatives of this exterior curvature angle. In this way we obtain
\ba\label{337}
\frac{\partial \omega_{opk}^A}{\partial l_{ij}} -\frac{\partial \omega_{opk}^B}{\partial l_{ij}} &=&
 (-1)^{s^A_i + s^A_j + s^A_o + s^A_p + s^A_k}\,  \frac{l_{ij} A_{opk}} {12}\, \frac{V_{\bar{o}} V_{\bar{p}} V_{\bar{k}} V_{\bar{i}} V_{\bar{j}}}{\prod_p V_{\bar{p}}}  
\ea
for all the boundary and bulk angles.  

To finally arrive at the (reduced) Hessian, we have to multiply this result (\ref{337}) with the area derivatives as in (\ref{bianeed}). This allows us to express the difference of the (reduced) Hessians in the $A$ and $B$ configurations as 
\begin{equation} \label{338}
H^{A}_{(op),(mn)} - H^{B}_{(op),(mn)}    \;=\; \underbrace{(-1)^{s^A_o + s^A_p + s^A_m +s^A_n}}_{=(-1)^{s^B_i+ \ldots + s^B_p}} \frac{l_{op} l_{mn}}{96} \frac{V_{\bar{o}} V_{\bar{p}} V_{\bar{m}} V_{\bar{n}}}{\prod_l V_{\bar{l}}} \underbrace{\sum_{k \neq o,p} \underbrace{(-1)^{s^A_k + 1}}_{=(-1)^{s^B_k}} F_{op;k} V_{\bar{k}}}_{:=D^A} 
\end{equation}


Also here $D$ does not depend on the choice of indices $o,p$ in (\ref{338}). Note that $D^A = - D^B$. Unless $D^A=D^B=0$, the (quadratic) action of the linearized theory is different in the $A$ and the $B$ configuration. This equality $D^A=D^B=0$ does not hold on general (flat) background configurations, but might hold in very symmetric cases.


 Furthermore, a measure of the form
\begin{equation}
\mu(l) = \frac{\prod_e \frac{l_e}{\sqrt{192 \pi}}}{\sqrt{\prod_l V_{\bar{l}}}}
\end{equation}
is only invariant under the $3-3$ move in the case that
\begin{equation} \label{eq:meas_inv_3-3}
V_{\bar{0}} V_{\bar{1}} V_{\bar{2}} = V_{\bar{3}} V_{\bar{4}} V_{\bar{5}}  \q .
\end{equation}
Again, this equality does  not hold for generic cases.

\section{Summary for 4D gravity} \label{sec:summary_4d}

4D classical Regge calculus is invariant under the $4-2$ and $5-1$ moves, but not under the $3-3$ moves. Our calculations provided the evidence for the linearized theory, in particular isolating the invariance breaking term for the $3-3$ move.

But this invariance behavior also holds for the full theory: there is always a flat solution to the equation of motions associated to the $4-2$ and $5-1$ Pachner moves. Hence the contribution from the bulk to the Hamilton-Jacobi function is vanishing. The Hamilton-Jacobi function is therefore just given by the boundary terms, which do not change under the $4-2$ and $5-1$ moves.

Concerning the quantum theory for linearized Regge Calculus in 4D, we define the path integral for general triangulations as
\begin{equation} \label{eq:path_int_4d}
P :=  \int    \frac{\prod_{e} \frac{l_e}{\sqrt{192 \pi}}}{\prod_{\Delta} \sqrt{V_{\Delta}}}\prod_{e \subset \text{bulk}} d\lambda_e \, \exp \left \{- \frac{1}{2} H_{e,e'} \lambda_e \lambda_{e'} \right\} \q .
\end{equation}
$l_e$ denotes the length of edge $e$, $V_\Delta$ the volume of 4--simplex $\Delta$, $\lambda_e$ is the edge length perturbation of edge $e$ and $H_{e,e'}$ is the $e$-$e'$ matrix element of the Hessian matrix of the Regge action. 

In the previous section we have shown that (\ref{eq:path_int_4d}) is invariant -- modulo the factor $D$ -- under $4-2$ and $5-1$ Pachner moves (using the gauge fixing conventions discussed above), but in general not under the $3-3$ Pachner move. The non-invariance under $3-3$ moves is already present in the classical theory and should be overcome by constructing a perfect discretization \cite{bahrdittrich2,bahrdittrich1,song,steinhaus,bahrreview}.

It might be possible to implement  a full invariance of the path integral under either the $4-2$ or the $5-1$ move, that is  by including the factor $\sqrt{D}$ into the measure. For the $4-2$ moves one would need to associate a corresponding factor to the edges of the triangulation, for the $5-1$ move rather to the vertices. (Alternatively, one would have to change the gauge fixing procedure for the $5-1$ move, i.e. the factor associated to the gauge orbit, but this seems to be rather unnatural.)  Still there are several open questions left to address, as how to generalize the definition of the $D$ factors to more complicated triangulations ( the bulk edges in the Pachner moves are always shared by four triangles) and how the $D$ factors associated to boundary edges or vertices will interfere. Furthermore, the factor $D$ is slightly non-local, but its actual form might be due to the linearized theory.

Here it might be helpful to reconsider the topological $BF$ theory, from which gravity can be obtained by implementing (simplicity) constraints. This is the route followed by spin foams. The advantage of this approach is, that a triangulation invariant path integral can be constructed for $BF$ theory. To apply this to Regge calculus one would need a formulation based on the same geometric variables as used in 4D $BF$ theory. Such a formulation is provided by area-angle Regge calculus \cite{area-angle}. The corresponding action can also be split into a piece describing a topological theory and constraints acting in the same way as the simplicity constraints. Studying this action might help to construct a triangulation independent quantum theory describing flat space dynamics. For other work in this direction, related to $BF$ theory see \cite{valentin_areaangle, valentin_ham,aristide}.


Despite all these subtleties and drawbacks, the simple form of the Hessian matrix for all Pachner moves and its similar form to the 3D case are remarkable. Therefore it will be very interesting to compare our results to spin foam asymptotics and possibly help to fix measure ambiguities (by requiring invariance under Pachner moves) there.

%

\section{Discussion} \label{sec:discussion}

In this work we provided  extensive analytical calculations for linearized Regge calculus, very much enlightening the structure of the theory. In particular we obtained the linearized Regge actions associated to all the Pachner moves in 3D and 4D, explicitly showing that the Regge action\footnote{This invariance result holds also for the full theory.} is invariant under all Pachner moves, with the exception of the $3-3$ move. We isolated the gauge symmetries and the conformal factor problem, which both are potential sources for divergencies.  Amazingly the structure of the linearized Regge actions associated to the Pachner moves lead in all cases to a very transparent factorizing structure, similar in 3D and in 4D. These formulae might be also helpful in other contexts, for instance in a canonical formulation of Regge calculus \cite{philipp} or in numerical larger scale calculations.

Furthermore we proposed a dynamical principle to fix the measure for Regge calculus, namely to consider the behavior of the theory under Pachner moves. Restricting to a local ansatz as in (\ref{intro2}) this fixes the measure uniquely.  Indeed the invariance under change of triangulation is related to an implementation of diffeomorphism symmetry \cite{dittrichreview,bahrdittrich2,steinhaus,bahrreview}.  This condition can therefore be understood as requiring an anomaly--free measure, which can be expected to be unique. A simple reason for this is that for a theory completely invariant under changes of the triangulation, there is also no (bulk) discretization scale. That is the only discretization scale is provided by the boundary data. For compact manifolds without boundary the continuum limit is even trivial, as such a limit would be obtained via a refinement of the triangulation \cite{dittinvariance}. In other words a triangulation independent path integral provides already  the continuum result.

This is the reason why we cannot expect to obtain a fully triangulation independent local theory in 4D. However one might ask for invariance of the quantum theory under the same set of local triangulation changes under which the local classical theory is invariant. Such a set can be understood as trivial subdivisions of the triangulation, as the associated equations of motions lead to flat space--time. The question therefore is, whether one can define a topological sub--sector of the theory \cite{dittrichryan,valentin_ham}, which would provide a quantum description of flat space dynamics, see also \cite{aristide}. Such an invariance of the theory under trivial subdivisions seems also be crucial to realize scenarios as proposed in \cite{dittinvariance}, i.\ e.\ the convergence of the theory to a topological sector, under refinement.

 We found a path integral measure for linearized Regge calculus, which provides such an invariance, in 4D modulo a factor, which features a certain non--local structure.  In 3D we found an exactly invariant measure, which also coincides with the asymptotics of the (triangulation independent) Ponzano Regge model. This is quite astonishing, as we performed a calculation in the linearized theory.  Furthermore the Ponzano Regge model includes in addition also a sum over orientations, which we do not consider here. The question arises, whether this result can be extended to the full non-linear theory and shed light on the problem, whether to include a sum over orientations into a quantum gravity path integral or not \cite{orientations,carlo_private}.

 The factor appearing in 4D, disturbing invariance at least under the $5-1$ and $4-2$ moves, is related to a transformation from area to length variables. It might therefore be helpful, in order to further enlighten this issue, to consider area-angle Regge calculus \cite{area-angle,newregge}. This formulation allows a split into a topological theory, which would be triangulation independent, and constraints. Another possibility to obtain path integral measures is a derivation from the canonical theory, which has recently became available for Regge calculus \cite{philipp}. Indeed the path integral measure is important to obtain correlation functions, which are annihilated by the Hamiltonian constraints \cite{steinhaus}.

A fully triangulation independent theory can be constructed via the method of perfect discretizations \cite{bahrdittrich1,bahrdittrich2,song,steinhaus}, which is based on a Wilsonian renormalization flow. This has the advantage of providing at the same time informations on the continuum limit of the theory. Here two different strategies can be thought of. One is based on local considerations, namely to study the behavior of a given theory under local refinements, e.\ g.\ Pachner moves, see for instance \cite{etera_valentin} for related studies in (topological) spin foam models.  This can result in recursion relations, whose fixed points provide the continuum limit (and perfect discretization) of the path integral, see \cite{steinhaus} for an example in 1D. Another strategy is to extract the large scale behavior, which might depend on the choice of measure \cite{lollreview,hambernumerics}. First steps towards extracting large scale behavior of (simplified) spin foam models via real space renormalization can be found in \cite{franke}.

\section*{Acknowledgements}

We thank Benjamin Bahr for providing us with his result concerning the invariance of classical Regge calculus under the $3-3$ pachner move. Furthermore we express our gratitude to Wojciech Kaminski for insights crucial in Appendix B. We also thank Aristide Baratin, Frank Hellmann and Ruth Williams for discussions.

\begin{appendix}
\section{Euclidean integration measure} \label{app:eucl_int}

The usual Lebesgue measure of $D$-dimensional Euclidean space can be rewritten with respect to the edge lengths of a (non-degenerate) $D$-simplex \cite{Baratin:2006yu}. 

Assume $D+1$ vertices making up a $D$-simplex embedded in $\mathbb{R}^D$, their positions given by $\{\vec{x}_i\}_{i=0,\ldots,D} $, such that the $D$-simplex is not degenerate, i.e. its $D$-volume is non-vanishing. Next, we define the position of the vertices of the $D$-simplex with respect to one of its vertices by defining $\vec{l}_i := \vec{x}_i - \vec{x}_0$. Since the $D$-simplex is non-degenerate, the set of vectors $\{\vec{l}_i \}_{i=1,\ldots,D}$ form a (non-orthonormal) basis of $\mathbb{R}^D$, where the lengths of the vectors $\vec{l}_i$ give the edge lengths $l_{0i}$ of the $D$-simplex. To write the Lebesgue measure in these coordinates, one has to compute the Jacobian of the linear function which maps the orthonormal basis $\{\vec{e}_i\}$ to $\{\vec{l}_i\}$. To simplify notation, we will denote $\vec{y}:= \vec{x}_0$.
\begin{equation} \label{eq:appa:measure}
\prod_{i=1}^D l_{0i} \, dl_{0i} = \prod_{i=1}^D d\left( \frac{l_{0i}^2}{2}\right) = \prod_{i=1}^D d \left(\frac{(\vec{y} - \vec{x}_i)^2}{2} \right) = \prod_{i=1}^D dy_i \left|\det \left[\frac{\partial}{\partial y^j} \frac{(\vec{y} - \vec{x}_i)^2}{2} \right] \right|
\end{equation}
where the determinant in the last term is the Jacobian of the coordinate transformation. For the matrix elements of the Jacobian one obtains:
\begin{equation} \label{eq:appa:jac}
\frac{\partial}{\partial y^j} \frac{(\vec{y}-\vec{x}_i)^2}{2} = \vec{e}_j \cdot (\vec{y} - \vec{x}_i)\q .
\end{equation}
Given \eqref{eq:appa:jac}, the Jacobian can be rewritten in terms of the volume of the $D$-simplex:
\begin{equation} \label{eq:appa:volume}
\left| \det \left(\vec{e}_j \cdot (\vec{y} - \vec{x}_i) \right) \right| = \sqrt{ \det \left( (\vec{y} -\vec{x}_j) \cdot (\vec{y} - \vec{x}_i) \right) } = D! \, V \q .
\end{equation}
Using (\ref{eq:appa:jac},\ref{eq:appa:volume}) in (\ref{eq:appa:measure}) gives:
\begin{eqnarray}
\prod_{i=1}^D l_{0i} \, dl_{0i} \;&=&\; \prod_{i=1}^D dy_i \; D! \, V \\
\implies \prod_{i=1}^D d y_i \;& = &\; \frac{\prod_{i=1}^D l_{0i}\, dl_{0i}}{D! \, V} \q . \label{eq:appa:final}
\end{eqnarray}
From (\ref{eq:appa:final}) one obtains (\ref{eq:eucl_m_3d}) in 3D and (\ref{eq:eucl_m_4d}) in 4D.

\section{Determinant formula for $D$}

In the calculations of the Hessian matrix in 4D, see section \ref{sec:Hessian_4d}, we have encountered the factors $D_{op}$ (see for instance (\ref{bianeed})) which are slightly non-local. In this section we will present a different way to compute these factors in terms of a determinant of a matrix which will additionally allow us to show that (some) factors $D_{op}$ are equal without using that the Hessian matrix is symmetric.

Consider six vertices embedded in $\mathbb{R}^4$ making up a triangulation which can be modified by one of the Pachner moves discussed in section \ref{sec:Pachner_4d}. We define the position vectors of the vertices with respect to vertex $(0)$, which we place in the origin of the coordinate system for simplicity. Hence the position vector of vertex $(i)$ is defined by $\vec{x}^i\equiv \vec{x}^i - \vec{x}^0$, its components are denoted by $x^i_a$ with $a=1,2,3,4$. Given this definition, consider the following determinant:
\begin{equation}
\det \left[ 
\begin{array}{c c c c}
\;x^1_0 \; & \; x^2_0 \; & \ldots & \; x^5_0 \; \\
x^1_1 & x^2_1 & \ldots & \\
\vdots & &\ddots & \\
x^1_4 & x^2_4 & \ldots & x^5_4 \\
\;(\vec{x}^1)^2 \; & \; (\vec{x}^2)^2 \;& \;\ldots \; & \;(\vec{x}^5)^2 \;
\end{array}
\right] \q .
\end{equation}
We will show that this determinant is proportional to the factors $D_{0i}$, for $i=1,\ldots,5$.

The determinant of a matrix remains unchanged if a scalar multiple of one of its rows / columns is added to one of its rows / columns respectively. Hence we will subtract $x^i_a$ times the $a$th row from the last for $1 \leq a<5$. Then one obtains for instance for $i=1$:
\begin{equation} \label{eq:determinantforD}
\det \left[ 
\begin{array}{c c c c} 
\;x^1_1 \; & \; x^2_1 \; & \ldots & \; x^5_1 \; \\
x^1_2 & x^2_2 & \ldots & \\
\vdots & &\ddots & \\
x^1_4 & x^2_4 & \ldots & x^5_4 \\
\;0 \; & \; (\vec{x}^2)^2 - \vec{x}^2 \cdot \vec{x}^1 \;& \;\ldots \; & \;(\vec{x}^5)^2 -\vec{x}^5 \cdot \vec{x}^1 \;
\end{array}
\right] \q .
\end{equation}
The terms in the last row can be rewritten as
\begin{equation}
(\vec{x}^k)^2 - \vec{x}^k \cdot \vec{x}^1 = \frac{1}{2} (\underbrace{(\vec{x}^k)^2 - (\vec{x}^1)^2+ (\vec{x}^k - \vec{x}^1)^2}_{=l_{0k}^2 - l_{01}^2 + l_{1k}^2} ) = \frac{1}{2} F_{01;k} \q ,
\end{equation}
where $F_{01;k}$ is the same factor as in (\ref{eq:der_area}). Expanding (\ref{eq:determinantforD}) with respect to the last row, the subdeterminants of the matrix correspond to (oriented) volumes of 4-simplices (see (\ref{eq:appa:volume})). Hence one obtains:
\begin{equation} \label{eq:det_for_D}
\underbrace{\det[\ldots]}_{(\ref{eq:determinantforD})} \; = \; 4! \sum_{i=2}^5 (-1)^{i-1} \, s_i \, F_{01;i} \, V_{\bar{i}} \; \propto \; D_{01} \q 
\end{equation}
where $s_i$ are appropriate sign factors taking into account the orientation of the volumes. 
By analogous considerations, one computes the other factors $D_{0j}$, $j=2,3,4,5$, from the same determinant, which implies that these factors are all equal (although the symmetry of the Hessian matrix has not been used). The signs $s_i$ depend on the relative orientation of the vectors $\vec{x}_i$ of the respective 4-simplex and hence they depend on the geometry under discussion. We would like to demonstrate that explicitly with the example of the $4-2$ move.

Consider the $D_{01}$ for the $4-2$ move, see (\ref{bianeed}). Four of the five edge vectors $\vec{x}_i$ of the vertices, defined with respect to the vertex $(0)$, form a basis of $\mathbb{R}^4$ since the 4-simplices are not degenerated. In the following we assume that $\vec{x}_2, \ldots, \vec{x}_5$ is a positively oriented orthonormal basis. Hence, we can rewrite the determinant (\ref{eq:determinantforD}) as
\begin{equation} \label{eq:sign_1}
\det \left[ 
\begin{array}{c c c c c} 
\;\sum_{i=2}^5 v_i \vec{x}_i \; & \; \vec{x}_2 \; & \; \vec{x}_3 \; & \; \vec{x}_4 \; & \; \vec{x}_5 \; \\
\;0 \; & \; F_{01;2} \;& \; F_{01;3} \; & \;F_{01;4} \; & \; F_{01;5} \;
\end{array}
\right] \q 
\end{equation}
where the $v_i$ are the coefficients of $\vec{x}_1$ in the basis formed by $\vec{x}_2,\ldots, \vec{x}_5$. Note that the configuration of the $4 - 2$ move can be chosen such that $v_i>0$.
If (\ref{eq:sign_1}) is expanded with respect to the last row, the submatrices are of the following form:
\begin{equation}
\det\left[\;\sum_{i=2}^5 v_i \vec{x}_i \; \;\vec{x}_j \; \; \vec{x}_k \; \; \vec{x}_l \; \right] = \epsilon_{ijkl}\, v_i \q ,
\end{equation}
where $\epsilon_{ijkl}$ is the Levi-Cevita symbol in 4D. Hence the signs $s_i$ in (\ref{eq:det_for_D}) are alternating, which, together with the alternating signs $(-1)^{i-1}$ verifies formula (\ref{eq:det_for_D}).
\end{appendix}

\end{document}